\documentclass[prl,aps,showpacs,twocolumn,
amsmath,amssymb,superscriptaddress]{revtex4-2}

\usepackage{graphicx}
\usepackage{dcolumn}
\usepackage{bm}
\usepackage{amsmath,amssymb,amsfonts}

\begin{document}

\preprint{APS/123-QED}

\title{Direct Measurement of Helicoid Surface States in RhSi using Nonlinear Optics}

\author{Dylan Rees}
\thanks{These two authors contributed equally}
\affiliation{Department of Physics, University of California, Berkeley, Berkeley, CA 94720, USA.}
\affiliation{Materials Science Division, Lawrence Berkeley National Laboratory, Berkeley, CA 94720, USA.}
\author{Baozhu Lu}
\thanks{These two authors contributed equally}
\affiliation{Department of Physics, Temple University, Philadelphia, PA 19122, USA}
\author{Yue Sun}
\affiliation{Materials Science Division, Lawrence Berkeley National Laboratory, Berkeley, CA 94720, USA.}
\affiliation{Department of Chemistry, University of California, Berkeley, Berkeley, CA 94720, USA.}
\author{Kaustuv Manna}
\affiliation{Max Planck Institute for Chemical Physics of Solids, Dresden D-01187, Germany}
\affiliation{Department of Physics, Indian Institute of Technology Delhi, New Delhi 110016, India}
\author{R\"ustem \"Ozg\"ur}
\affiliation{Department of Materials Science and Engineering, University of California, Berkeley, CA
94720, USA.}
\author{Sujan Subedi}
\affiliation{Department of Physics, Temple University, Philadelphia, PA 19122, USA}
\author{Claudia Felser}
\affiliation{Max Planck Institute for Chemical Physics of Solids, Dresden D-01187, Germany}
\author{J. Orenstein}
\email{jworenstein@lbl.gov}
\affiliation{Department of Physics, University of California, Berkeley, Berkeley, CA 94720, USA.}
\affiliation{Materials Science Division, Lawrence Berkeley National Laboratory, Berkeley, CA 94720, USA.}
\author{Darius H. Torchinsky} \email{dtorchin@temple.edu}
\affiliation{Department of Physics, Temple University, Philadelphia, PA 19122, USA}

\date{\today}

\begin{abstract}
Despite the fundamental nature of the edge state in topological physics, direct measurement of electronic and optical properties of the Fermi arcs of topological semimetals has posed a significant experimental challenge, as their response is often overwhelmed by the metallic bulk. However, laser-driven currents carried by surface and bulk states can propagate in different directions in nonsymmorphic crystals, allowing for the two components to be easily separated. Motivated by a recent theoretical prediction \cite{chang20}, we have measured the linear and circular photogalvanic effect currents deriving from the Fermi arcs of the nonsymmorphic, chiral Weyl semimetal RhSi over the $0.45 - 1.1$~eV incident photon energy range. Our data are in good agreement with the predicted magnitude of the circular photogalvanic effect as a function of photon energy, although the direction of the surface photocurrent departed  from the theoretical expectation over the energy range studied. Surface currents arising from the linear photogalvanic effect were observed as well, with the unexpected result that only two of the six allowed tensor element were required to describe the measurements, suggesting an approximate emergent mirror symmetry inconsistent with the space group of the crystal.
\end{abstract}



\maketitle


A universal property of topological matter is the existence of a protected edge state, e.g., the current-carrying edge state of the quantum Hall effect~\cite{Hasan10,Qi11} or the spin-momentum locked surface states of bulk topological insulators~\cite{Fu07,Hsieh08}. In topological Weyl semimetals, which host emergent massless, chiral charge carriers called Weyl fermions, the topological edge state comprises open Fermi surface arcs formed of helicoidally dispersing, spin-momentum locked quasiparticles that are constrained to the sample surface~\cite{wan11, armitage18}. The arcs connect the projections of opposite chirality Weyl nodes, curving in complementary shapes on the 2D surface Brillouin zones on the opposite sides of the crystal. The existence of these states has been confirmed by ARPES \cite{xu15,lv15,xu15_2,belopolski16} and quasiparticle interference measurements \cite{inoue16} and have been shown to play a central role in quantum oscillations~\cite{potter2014,moll16}. However, despite a number of predictions focused on the role of the Fermi arcs in topological semimetal physics~\cite{jia16,shi2017,song2017,Mukherjee2019,ghosh2020,wawrzik20}, their transport and optical properties have largely remained hidden, as they are often dominated by bulk response functions. For example, experiments aimed at measuring the linear conductivity of surface states run into difficulties because of shorting by the metallic bulk that lies below. 


\begin{figure*}
\includegraphics{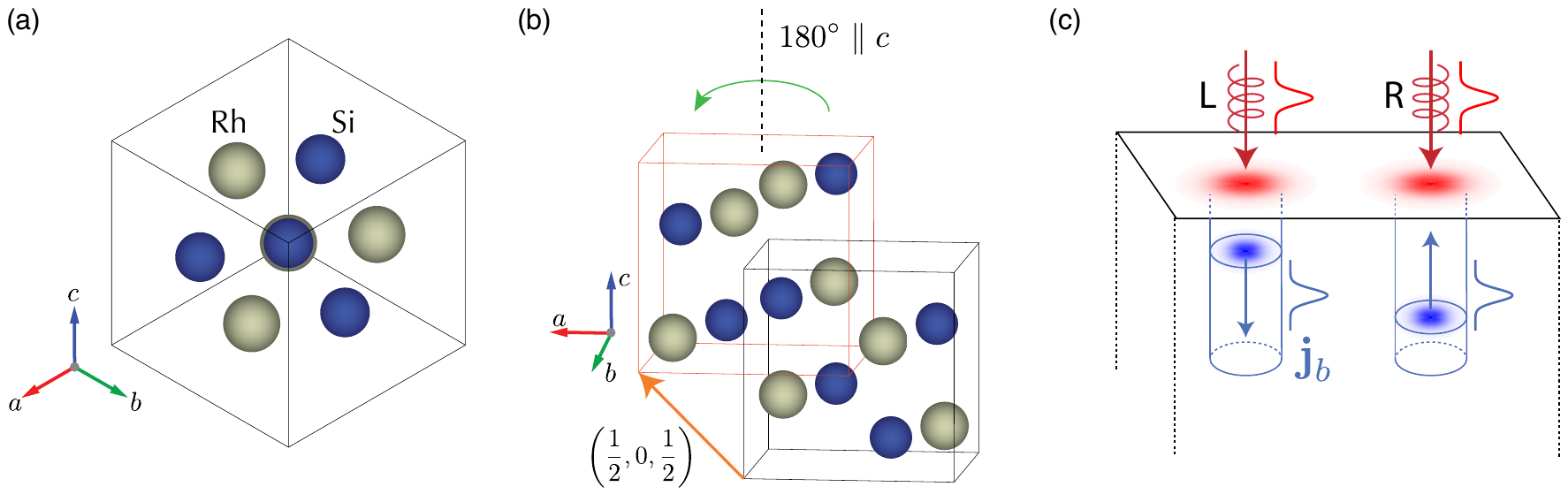}
\caption{\label{fig:1} (a) Unit cell displayed with the [111] direction pointing out of the page, showing the three-fold rotational symmetry of the crystal. (b) Extended RhSi structure, showing two alternate unit cells offset by $\left (1/2, 0,1/2 \right )$. When the unit cell marked by the orange frame is rotated $180^\circ$ about the $z$ axis, it is identical to the unit cell marked by the black frame, illustrating the two-fold screw symmetry. (c) When circularly polarized light is incident on RhSi, the bulk CPGE current will be directed perpendicular to the surface, with its sign determined by the incident light's handedness. \textit L and \textit R refer to left- and right-handed circular polarization and $\textbf j_b$ refers to the bulk CPGE current.}
\end{figure*}

\begin{figure}[b]
\includegraphics{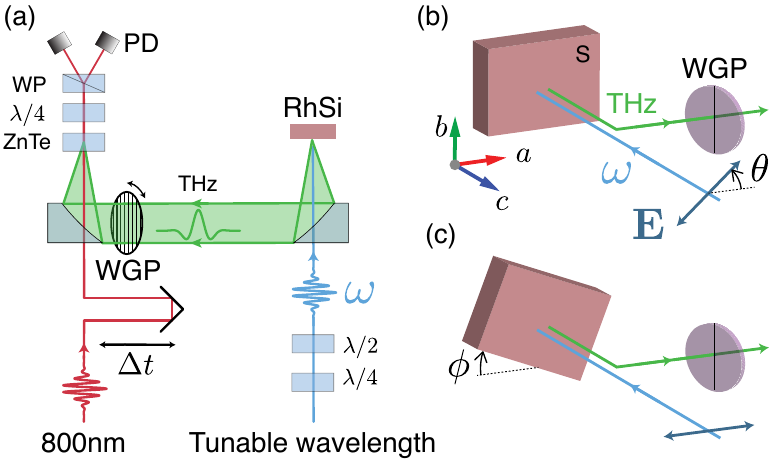}
\caption{\label{fig:2} (a) Schematic of experiment used to detect photogalvanic currents in RhSi via terahertz detection. Near infrared (NIR) light with tunable wavelength and polarization is focused onto the [001] RhSi surface at normal incidence. Terahertz radiation is collected and collimated using off-axis parabolic mirrors. It passes through a wire-grid polarizer before being focused onto a ZnTe crystal. Light with $\lambda=800$~nm and variable time delay $\Delta t$ copropagates through the ZnTe for electro-optical detection of the terahertz. PD, photodiode; WP, Wollaston prism; WGP wire grid polarizer; $\lambda/2$, half-wave plate; $\lambda/4$, quarter-wave plate.  (b) In one experimental configuration, the sample is kept fixed while the pump polarization is rotated by angle $\theta$. The sample axes are set such that [100] and [010] are horizontal and vertical in the lab frame respectively. (c) In the second configuration, the pump polarization is fixed at $\theta=0$ and the sample is rotated by and angle $\phi$.}
\end{figure}

In this work we demonstrate experimentally that the second-order nonlinear conductivity, which describes the strength and symmetry of the photogalvanic effects (PGEs), provides a means to selectively probe surface state electronic properties in Weyl semimetals. The PGEs are phenomena in which optical excitation generates a dc current that arises from intrinsic breaking of inversion symmetry, rather than applied bias voltage or inhomogeneous doping. A further defining property of PGEs is sensitivity of the direction of photocurrent, $\textbf J$, to the polarization state of the optical electric field $\textbf E$, as described by the phenomenological relation,
\begin{equation}
J_{i}=\gamma_{i j k}E_{j} E_{k}+i \beta_{i j}(\textbf E \times \textbf E^*)_{j}.
\end{equation}
The first term on the right-hand side of Eq. 1 describes a current generated by linearly polarized light (the LPGE) in terms of the polar tensor $\gamma_{ijk}$. The second term corresponds to a photocurrent whose direction reverses with reversal of the helicity of the photoexcitation. This circular PGE (CPGE) is proportional to the axial tensor $\beta_{ij}$. Both PGE response tensors are zero in the presence of inversion symmetry.


The CPGE has received particular attention in the Weyl semimetal RhSi and related topological semimetals, because they crystallize in structures in which all mirror symmetries are broken, forming a chiral (or handed) medium \cite{chang17,tang17,chang18,flicker18,le20,ni20,li19,cochran20,sanchez19}. In chiral Weyl semimetals, nodes with opposite topological charge need not be degenerate in energy allowing for one node to lie near the Fermi energy, $E_F$, while its oppositely charged partner may be well below \cite{dejuan17,martinez19,bradlyn16,sanchez19}. The breaking of degeneracy creates a photon energy window in which CPGE arises exclusively from the node near $E_F$, theoretically allowing a quantized CPGE to emerge with amplitude directly proportional to its Berry monopole charge \cite{dejuan17}. However, recent experiments with light incident on the (111) surface of RhSi and related isostructural compounds have shown that ideal quantization of the CPGE is disrupted by optical transitions between non-Weyl bands that lie within the quantization window~\cite{rees20,maulana20}.  Nevertheless, it was also shown that the polarization selection rules for both CPGE and LPGE observed on (111)  faithfully follow constraints imposed by the symmetry of the bulk \cite{rees20}.  As we show below, these constraints provide a route to selectively probe the Fermi arc surface states on the (001) surface.

The space group of RhSi (\#198) contains two operations: a 3-fold rotation about the [111] direction and a nonsymmorphic screw symmetry in which a 2-fold rotation about the $z$ axis is combined with a translation by $\left ( 1/2, 0, 1/2 \right )$ (Fig.~1(a-b)). In describing bulk response functions, where perfect translational symmetry is assumed, the screw operation imposes the same constraints on response tensors as 2-fold rotation. The combination of the 3 and 2-fold rotational symmetry greatly reduces the number of nonvanishing elements of the $\gamma_{ijk}$ and $\beta_{ij}$ tensors that describe the bulk PGE response. Only tensor elements $\gamma_{xyz}=\gamma_{yzx}=\gamma_{zxy}$ of the LPGE response are nonzero, and the CPGE tensor is purely diagonal with $\beta_{ij}=\beta \delta_{ij}$. Note that given the reduction of the CPGE tensor to a scalar, Eq. 1 predicts that the CPGE current flows parallel to the wavevector of excitation light, independent of the crystal orientation.

As mentioned above, previous studies with light incident on the (111) surface verified the symmetry-based predictions for the bulk response functions \cite{rees20}. Specifically, the CPGE signal was below the noise level at normal incidence, consistent with the prediction that it flows parallel to the optical wavevector and therefore does not radiate in the direction of specular reflection (Fig.~1(c)). As further confirmation, THz radiation from CPGE current two orders of magnitude above the noise level was observed when the angle of incidence was set $45^\circ$ from the normal direction, where the bulk symmetry and measurement geometry imply a radiating CPGE current parallel to the surface.   

The experiments described below were stimulated by the prediction that the photogalvanic response to light normally incident on the (001) surface would be qualitatively different than (111), directly revealing the presence of topologically protected surface states through the observation of a surface current \cite{chang20}. Note that for (001) the symmetry of the bulk predicts that LPGE as well as CPGE current flows normal to the surface (see SI), in which case no radiation from PGEs is expected, as with CPGE on the (111) surface. The crucial ingredient leading to the prediction of PGE currents parallel to the (001) is the presence of a screw symmetry in the space group. Truncation of the crystal at (001) disrupts the translational component of the screw operation and violates the effective 2-fold symmetry. Consequently there is no operation, other than the identity, that transforms the (001) surface to itself and all tensor elements disallowed by bulk symmetry become allowed for surface-localized electronic states. In particular the six elements with only $x$ and $y$ indices (i.e, $\gamma_{xxx}$, $\gamma_{xxy}$, $\gamma_{xyy}$, $\gamma_{yxx}$, $\gamma_{yxy}$ and $\gamma_{yyy}$)  are not forbidden, allowing for in-plane photocurrent and specular THz radiation to be generated by light at normal incidence.

The apparatus used to observe short-lived surface-currents via their THz radiation is shown in Fig.~2(a). The excitation source was an optical parametric amplifier pumped by an amplified Ti:Sapphire laser, producing wavelength tunable pulses from 1150-2600 nm (0.48-1.1 eV) and pulse duration $\approx$100 fs. In-plane photogalvanic currents radiated a THz pulse into free space that was focused onto a ZnTe crystal for time-resolved electro-optic sampling of the THz transient (whereas the radiation due to through-plane bulk photocurrents did not emerge from the sample) \cite{rees20}. 

Figs. 2(b) and 2(c) show the experimental configurations used to measure the direction of the PGE currents for different polarization states of the normally incident radiation. The incident light was chosen to be either left or right circularly polarized, or linearly polarized with the plane of polarization rotatable through an angle $\theta$ (Fig.~2(b). In addition, the sample was also rotated about the optic axis by an angle $\phi$ (Fig.~2(c)). The crystal axes were determined by Laue diffraction and the sample rotation stage was initialized such that at $\phi=0$ (100) and (010) crystal axes are horizontal and vertical in the laboratory reference frame, respectively (see SI). Further information on the (001) oriented RhSi samples used here can also be found in the SI.

\begin{figure}[b]
\includegraphics{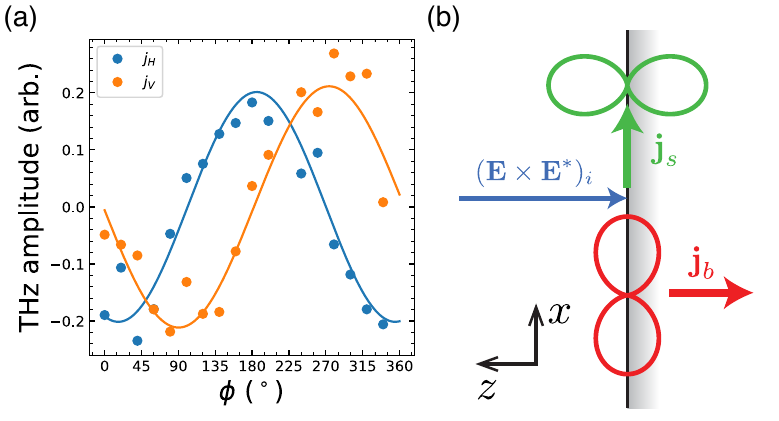}
\caption{\label{fig:3} (a) Amplitude of CPGE for horizontally and vertically polarized THz emission as a function of sample orientation $\phi$. (b) Schematic showing directions of bulk PGE ($\textbf j_b$, red) and surface PGE ($\textbf j_s$, green) with normally incident light on the 001 surface of RhSi with the resulting radiation patterns. In general $\textbf j_s$ has component in $x$ and $y$.}
\end{figure}

Figure 3(a) shows direct evidence for the generation of in-plane helicity-dependent photocurrent at normal incidence on the (001) surface. The THz amplitude plotted on the vertical axis is the difference in radiation generated by left and right circularly polarized light $\hbar\omega=0.8$ eV  and is thus a measure of the CPGE. The two plots show the dependence of the horizontal (H) and vertical (V) components of the CPGE amplitude on the angle of rotation, $\phi$, of the sample about the optic axis. The fact that CPGE is observable at normal incidence already suggests that in-plane photcurrent is generated. As the Fig. 3(b) illustrates, the dipole radiation pattern for normally directed photocurrent has a node at the angle of specular reflection from the surface and therefore does not directly generate THz radiation, although weaker radiation from multiple scattering is possible.  The proof that the observed radiation does indeed arise from an in-plane CPGE current is the dependence of the H and V components of the CPGE radiation on $\phi$. The solid lines in Fig. 3(a) are fits to $A \cos(\phi-\phi_0)$ and $A \sin(\phi-\phi_0)$, with $\phi_0\approx10^\circ$ for both components. This dependence of the CPGE amplitude on $\phi$ proves that as the sample rotates the CPGE current rotates as well, maintaining an angle $\phi_0$ with respect to the [100] direction . This behvavior is contrast to a normally directed CPGE current, which would be independent of $\phi$.

\begin{figure*}
\includegraphics{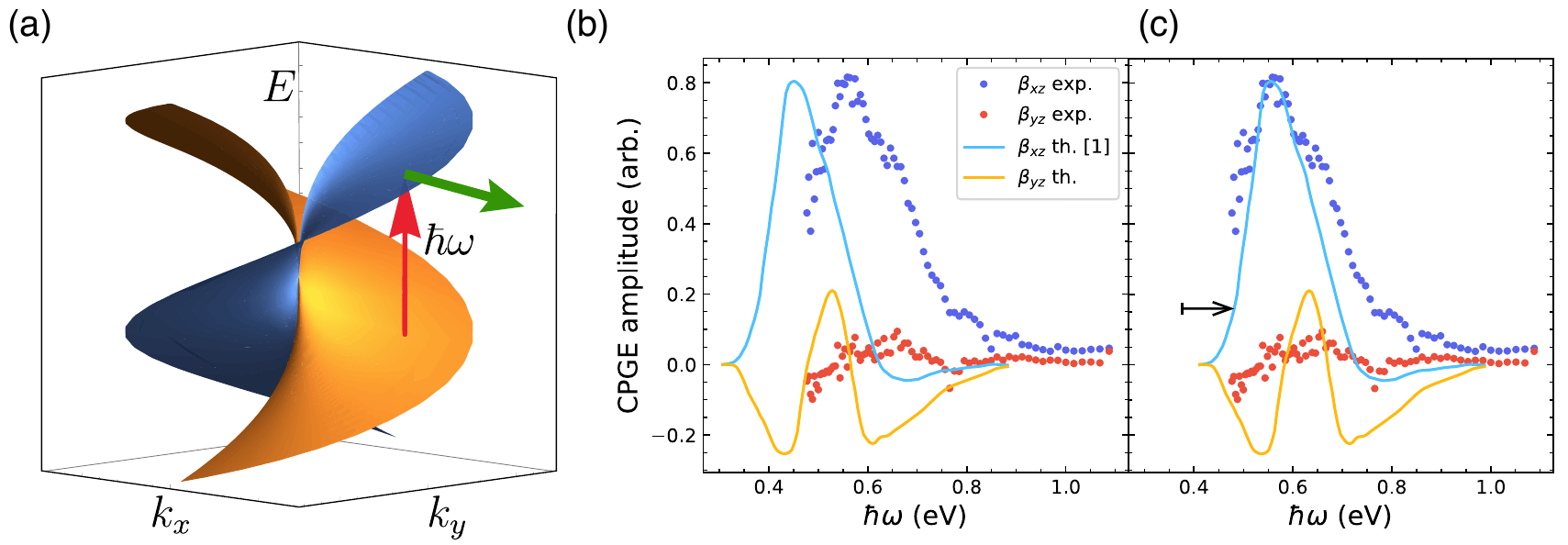}
\caption{\label{fig:4} (a) Schematic of surface helicoid bands including a photoexcitation of an electron at energy $\hbar\omega$ (red arrow) and the induced current (green arrow). (b) CPGE spectral data for $\beta_{xz}$ and $\beta_{yz}$ compared with theory from Ref.~\cite{chang20}. The second plot shows the data plotted with theory shifted by $0.1$~eV to demonstrate similarity.}
\end{figure*}

Having shown that a CPGE surface current is observed in violation of the restrictions placed by the symmetry of the bulk, we next tested the theoretical prediction for the dependence of CPGE amplitude and direction on $\hbar\omega$. The surface bands responsible for Fermi arcs in RhSi comprise two intertwined helicoids with opposite spin polarization, as illustrated schematically in Fig.~4(a) in a plot of energy vs. in-plane momentum \cite{fang16,sanchez19}. The helicity-dependent in-plane CPGE current arises from spin-flip optical transitions between the two helicoids, as indicated by the arrows in Fig.~4(a).

Fig.~4(b) compares the observed CPGE amplitude (closed circles) as a function of $\hbar\omega$ with the spectra theoretically predicted from Wannier functions derived from first principles calculations (solid lines)~\cite{chang20}.  The two curves correspond to the H and V components of the CPGE current, proportional to $\beta_{xz}$ and $\beta_{yz}$, respectively. Fig.~4(c) shows the result of shifting the predicted spectra to higher photon energy by $\sim0.1$~eV to highlight the correlation between theory and experiment. The comparison shows a striking overall correspondence after a shift in the energy scale. One main difference is that the experiment shows that although the direction of the CPGE current varies with photon energy, it remains much closer to [100] than the theoretical prediction.

To fully characterize the nonlinear response, we measured the response to linear polarization, i.e., LPGE, in addition to the CPGE. Although Ref.~\cite{chang20} did not provide theoretical predictions for the LPGE, the implication of that work is that since 2-fold rotational symmetry is broken at the surface, the six elements of $\gamma_{ijk}$ that contain only $x$ and $y$ indices, forbidden in the bulk response, become allowed at the (001) surface. This symmetry-based argument would then predict the existence of in-plane LPGE currents whose directions need not correlate or align with the cubic axes of the crystal. 

\begin{figure}[b]
\includegraphics{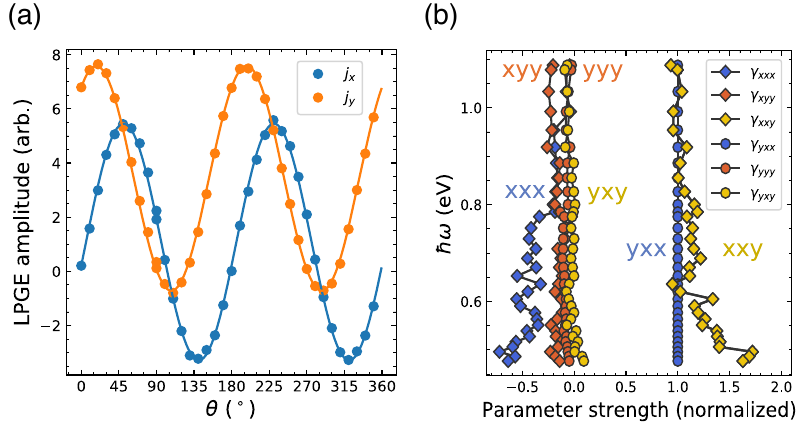}
\caption{\label{fig:5} (a) Terahertz amplitude along $x$ and $y$ as a function of linear pump polarization angle $\theta$. (b) Results of fitting data in (a) to general LPGE tensor $\gamma_{ijk}$.}
\end{figure}

As was the case with circularly polarized light, THz radiation was readily observed at normal incidence under photoexcitation with linearly polarized light. To determine the components of the LPGE tensor we resolved the THz amplitude into the H and V channels, varying the polarization angle of the pump beam while keeping the sample fixed. Fig.~5(a) shows the amplitude of the H and V components as a function of the angle of linear polarization, $\theta$. The solid lines are fits to $A \cos[2\theta-\theta_0]+B$. The six independently determined parameters, i.e., the amplitude of the cosine component, $A$, offset angle $\theta_0$, and offset amplitude $B$ for the H and V channels, are sufficient to determine the relative amplitude of all six elements of $\gamma_{ijk}$ that contribute to an in-plane current at normal incidence (see SI).  

Fig. 5(b) shows the relative amplitude of the six elements of $\gamma_{ijk}$ (normalized to $\gamma_{yxx}$) in the photon energy range from 0.5 to 1.2 eV. A striking feature of the spectra is that the response at photon energies above $\sim0.6$ eV  is dominated by two approximately equal components, $\gamma_{yxx}\approx\gamma_{xxy}$, with the other four close to zero, despite the fact that all six tensor components are in principle symmetry allowed. We note that the vanishing of components with an odd number of $x$ indices would suggest a mirror symmetry $x\rightarrow -x$ (see SI). A CPGE current directed along the $x$ axis would be consistent with this symmetry. While the components appear to approximately obey such a symmetry, we know of no mechanism which would enforce this. AFM measurements also revealed no patterns in surface topography that could affect the amplitude of photocurrents (see SI). 

In conclusion, we have measured the CPGE and LPGE response of Fermi arc surface states of the chiral Weyl semimetal RhSi over the energy range $0.45 - 1.1$~eV, confirming the prediction of a new path to selective probing of the topological surface states in Weyl semimetals. The measured CPGE spectrum matched the theoretical prediction to within a shift of the energy axis, while the angle of the photocurrent varied appreciably less than the theoretical prediction. LPGE measurements also probed the nonlinear response of Fermi arc surface states. An unanticipated result was that over a broad photon energy range only two elements of the nonlinear response tensor, $\gamma_{ijk}$ were required to fit the data, despite the six potentially nonzero elements expected to be allowed by the $C_1$ surface symmetry. This results presents a challenge to theory of surface states in Weyl semimetals. Finally, the measurement scheme demonstrated here provides a means to study way surface states of categories of topological matter through the use of nonlinear optical probes whose bulk response may be suppressed through symmetry.

{\bf Acknowledgements:}
We acknowledge Dan Parker, Joel Moore and Qimin Yan for useful conversations. {\bf Funding:} J.O. was supported by the Quantum Materials program, Director, Office of Science, Office of Basic Energy Sciences, Materials Sciences and Engineering Division, of the U.S. Department of Energy under Contract No. DE-AC02-05CH11231. J.O. received support for optical measurements from the Gordon and Betty Moore Foundation's EPiQS Initiative through Grant GBMF4537 to J.O. at UC Berkeley. The work at Temple University was funded by the National Science Foundation under Award No. NSF/DMR-1945222. K.M., and C.F. acknowledge the financial support from the European Research Council (ERC) Advanced Grant No. 742068 ”TOP-MAT”; European Union’s Horizon 2020 research and innovation program (Grant Nos. 824123 and 766566) and Deutsche Forschungsgemeinschaft (DFG) through SFB 1143. K. M. acknowledges the Max Planck Society for the funding support under Max Planck–India partner group project. R.\"O. acknowledges funding from SRC ACSENT program and Turkish Fulbright Commission.

\bibliography{apssamp}

\providecommand{\noopsort}[1]{}\providecommand{\singleletter}[1]{#1}%
\begin{thebibliography}{35}%
\makeatletter
\providecommand \@ifxundefined [1]{%
 \@ifx{#1\undefined}
}%
\providecommand \@ifnum [1]{%
 \ifnum #1\expandafter \@firstoftwo
 \else \expandafter \@secondoftwo
 \fi
}%
\providecommand \@ifx [1]{%
 \ifx #1\expandafter \@firstoftwo
 \else \expandafter \@secondoftwo
 \fi
}%
\providecommand \natexlab [1]{#1}%
\providecommand \enquote  [1]{``#1''}%
\providecommand \bibnamefont  [1]{#1}%
\providecommand \bibfnamefont [1]{#1}%
\providecommand \citenamefont [1]{#1}%
\providecommand \href@noop [0]{\@secondoftwo}%
\providecommand \href [0]{\begingroup \@sanitize@url \@href}%
\providecommand \@href[1]{\@@startlink{#1}\@@href}%
\providecommand \@@href[1]{\endgroup#1\@@endlink}%
\providecommand \@sanitize@url [0]{\catcode `\\12\catcode `\$12\catcode
  `\&12\catcode `\#12\catcode `\^12\catcode `\_12\catcode `\%12\relax}%
\providecommand \@@startlink[1]{}%
\providecommand \@@endlink[0]{}%
\providecommand \url  [0]{\begingroup\@sanitize@url \@url }%
\providecommand \@url [1]{\endgroup\@href {#1}{\urlprefix }}%
\providecommand \urlprefix  [0]{URL }%
\providecommand \Eprint [0]{\href }%
\providecommand \doibase [0]{http://dx.doi.org/}%
\providecommand \selectlanguage [0]{\@gobble}%
\providecommand \bibinfo  [0]{\@secondoftwo}%
\providecommand \bibfield  [0]{\@secondoftwo}%
\providecommand \translation [1]{[#1]}%
\providecommand \BibitemOpen [0]{}%
\providecommand \bibitemStop [0]{}%
\providecommand \bibitemNoStop [0]{.\EOS\space}%
\providecommand \EOS [0]{\spacefactor3000\relax}%
\providecommand \BibitemShut  [1]{\csname bibitem#1\endcsname}%
\let\auto@bib@innerbib\@empty
\bibitem [{\citenamefont {Chang}\ \emph {et~al.}(2020)\citenamefont {Chang},
  \citenamefont {Yin}, \citenamefont {Neupert}, \citenamefont {Sanchez},
  \citenamefont {Belopolski}, \citenamefont {Zhang}, \citenamefont {Cochran},
  \citenamefont {Ch\'eng}, \citenamefont {Hsu}, \citenamefont {Huang},
  \citenamefont {Lian}, \citenamefont {Xu}, \citenamefont {Lin},\ and\
  \citenamefont {Hasan}}]{chang20}%
  \BibitemOpen
  \bibfield  {author} {\bibinfo {author} {\bibfnamefont {G.}~\bibnamefont
  {Chang}}, \bibinfo {author} {\bibfnamefont {J.-X.}\ \bibnamefont {Yin}},
  \bibinfo {author} {\bibfnamefont {T.}~\bibnamefont {Neupert}}, \bibinfo
  {author} {\bibfnamefont {D.~S.}\ \bibnamefont {Sanchez}}, \bibinfo {author}
  {\bibfnamefont {I.}~\bibnamefont {Belopolski}}, \bibinfo {author}
  {\bibfnamefont {S.~S.}\ \bibnamefont {Zhang}}, \bibinfo {author}
  {\bibfnamefont {T.~A.}\ \bibnamefont {Cochran}}, \bibinfo {author}
  {\bibfnamefont {Z.~c. v. b.~a.}\ \bibnamefont {Ch\'eng}}, \bibinfo {author}
  {\bibfnamefont {M.-C.}\ \bibnamefont {Hsu}}, \bibinfo {author} {\bibfnamefont
  {S.-M.}\ \bibnamefont {Huang}}, \bibinfo {author} {\bibfnamefont
  {B.}~\bibnamefont {Lian}}, \bibinfo {author} {\bibfnamefont {S.-Y.}\
  \bibnamefont {Xu}}, \bibinfo {author} {\bibfnamefont {H.}~\bibnamefont
  {Lin}}, \ and\ \bibinfo {author} {\bibfnamefont {M.~Z.}\ \bibnamefont
  {Hasan}},\ }\href {\doibase 10.1103/PhysRevLett.124.166404} {\bibfield
  {journal} {\bibinfo  {journal} {Phys. Rev. Lett.}\ }\textbf {\bibinfo
  {volume} {124}},\ \bibinfo {pages} {166404} (\bibinfo {year}
  {2020})}\BibitemShut {NoStop}%
\bibitem [{\citenamefont {Hasan}\ and\ \citenamefont {Kane}(2010)}]{Hasan10}%
  \BibitemOpen
  \bibfield  {author} {\bibinfo {author} {\bibfnamefont {M.~Z.}\ \bibnamefont
  {Hasan}}\ and\ \bibinfo {author} {\bibfnamefont {C.~L.}\ \bibnamefont
  {Kane}},\ }\href {\doibase 10.1103/RevModPhys.82.3045} {\bibfield  {journal}
  {\bibinfo  {journal} {Rev. Mod. Phys.}\ }\textbf {\bibinfo {volume} {82}},\
  \bibinfo {pages} {3045} (\bibinfo {year} {2010})}\BibitemShut {NoStop}%
\bibitem [{\citenamefont {Qi}\ and\ \citenamefont {Zhang}(2011)}]{Qi11}%
  \BibitemOpen
  \bibfield  {author} {\bibinfo {author} {\bibfnamefont {X.-L.}\ \bibnamefont
  {Qi}}\ and\ \bibinfo {author} {\bibfnamefont {S.-C.}\ \bibnamefont {Zhang}},\
  }\href {\doibase 10.1103/RevModPhys.83.1057} {\bibfield  {journal} {\bibinfo
  {journal} {Rev. Mod. Phys.}\ }\textbf {\bibinfo {volume} {83}},\ \bibinfo
  {pages} {1057} (\bibinfo {year} {2011})}\BibitemShut {NoStop}%
\bibitem [{\citenamefont {Fu}\ \emph {et~al.}(2007)\citenamefont {Fu},
  \citenamefont {Kane},\ and\ \citenamefont {Mele}}]{Fu07}%
  \BibitemOpen
  \bibfield  {author} {\bibinfo {author} {\bibfnamefont {L.}~\bibnamefont
  {Fu}}, \bibinfo {author} {\bibfnamefont {C.~L.}\ \bibnamefont {Kane}}, \ and\
  \bibinfo {author} {\bibfnamefont {E.~J.}\ \bibnamefont {Mele}},\ }\href
  {\doibase 10.1103/PhysRevLett.98.106803} {\bibfield  {journal} {\bibinfo
  {journal} {Phys. Rev. Lett.}\ }\textbf {\bibinfo {volume} {98}},\ \bibinfo
  {pages} {106803} (\bibinfo {year} {2007})}\BibitemShut {NoStop}%
\bibitem [{\citenamefont {Hsieh}\ \emph {et~al.}(2008)\citenamefont {Hsieh},
  \citenamefont {Qian}, \citenamefont {Wray}, \citenamefont {Xia},
  \citenamefont {Hor}, \citenamefont {Cava},\ and\ \citenamefont
  {Hasan}}]{Hsieh08}%
  \BibitemOpen
  \bibfield  {author} {\bibinfo {author} {\bibfnamefont {D.}~\bibnamefont
  {Hsieh}}, \bibinfo {author} {\bibfnamefont {D.}~\bibnamefont {Qian}},
  \bibinfo {author} {\bibfnamefont {L.}~\bibnamefont {Wray}}, \bibinfo {author}
  {\bibfnamefont {Y.}~\bibnamefont {Xia}}, \bibinfo {author} {\bibfnamefont
  {Y.~S.}\ \bibnamefont {Hor}}, \bibinfo {author} {\bibfnamefont {R.~J.}\
  \bibnamefont {Cava}}, \ and\ \bibinfo {author} {\bibfnamefont {M.~Z.}\
  \bibnamefont {Hasan}},\ }\href {\doibase 10.1038/nature06843} {\bibfield
  {journal} {\bibinfo  {journal} {Nature}\ }\textbf {\bibinfo {volume} {452}},\
  \bibinfo {pages} {970} (\bibinfo {year} {2008})}\BibitemShut {NoStop}%
\bibitem [{\citenamefont {Wan}\ \emph {et~al.}(2011)\citenamefont {Wan},
  \citenamefont {Turner}, \citenamefont {Vishwanath},\ and\ \citenamefont
  {Savrasov}}]{wan11}%
  \BibitemOpen
  \bibfield  {author} {\bibinfo {author} {\bibfnamefont {X.}~\bibnamefont
  {Wan}}, \bibinfo {author} {\bibfnamefont {A.~M.}\ \bibnamefont {Turner}},
  \bibinfo {author} {\bibfnamefont {A.}~\bibnamefont {Vishwanath}}, \ and\
  \bibinfo {author} {\bibfnamefont {S.~Y.}\ \bibnamefont {Savrasov}},\ }\href
  {\doibase 10.1103/PhysRevB.83.205101} {\bibfield  {journal} {\bibinfo
  {journal} {Phys. Rev. B}\ }\textbf {\bibinfo {volume} {83}},\ \bibinfo
  {pages} {205101} (\bibinfo {year} {2011})}\BibitemShut {NoStop}%
\bibitem [{\citenamefont {Armitage}\ \emph {et~al.}(2018)\citenamefont
  {Armitage}, \citenamefont {Mele},\ and\ \citenamefont
  {Vishwanath}}]{armitage18}%
  \BibitemOpen
  \bibfield  {author} {\bibinfo {author} {\bibfnamefont {N.~P.}\ \bibnamefont
  {Armitage}}, \bibinfo {author} {\bibfnamefont {E.~J.}\ \bibnamefont {Mele}},
  \ and\ \bibinfo {author} {\bibfnamefont {A.}~\bibnamefont {Vishwanath}},\
  }\href {\doibase 10.1103/RevModPhys.90.015001} {\bibfield  {journal}
  {\bibinfo  {journal} {Rev. Mod. Phys.}\ }\textbf {\bibinfo {volume} {90}},\
  \bibinfo {pages} {015001} (\bibinfo {year} {2018})}\BibitemShut {NoStop}%
\bibitem [{\citenamefont {Xu}\ \emph {et~al.}(2015{\natexlab{a}})\citenamefont
  {Xu}, \citenamefont {Liu}, \citenamefont {Kushwaha}, \citenamefont {Sankar},
  \citenamefont {Krizan}, \citenamefont {Belopolski}, \citenamefont {Neupane},
  \citenamefont {Bian}, \citenamefont {Alidoust}, \citenamefont {Chang},
  \citenamefont {Jeng}, \citenamefont {Huang}, \citenamefont {Tsai},
  \citenamefont {Lin}, \citenamefont {Shibayev}, \citenamefont {Chou},
  \citenamefont {Cava},\ and\ \citenamefont {Hasan}}]{xu15}%
  \BibitemOpen
  \bibfield  {author} {\bibinfo {author} {\bibfnamefont {S.-Y.}\ \bibnamefont
  {Xu}}, \bibinfo {author} {\bibfnamefont {C.}~\bibnamefont {Liu}}, \bibinfo
  {author} {\bibfnamefont {S.~K.}\ \bibnamefont {Kushwaha}}, \bibinfo {author}
  {\bibfnamefont {R.}~\bibnamefont {Sankar}}, \bibinfo {author} {\bibfnamefont
  {J.~W.}\ \bibnamefont {Krizan}}, \bibinfo {author} {\bibfnamefont
  {I.}~\bibnamefont {Belopolski}}, \bibinfo {author} {\bibfnamefont
  {M.}~\bibnamefont {Neupane}}, \bibinfo {author} {\bibfnamefont
  {G.}~\bibnamefont {Bian}}, \bibinfo {author} {\bibfnamefont {N.}~\bibnamefont
  {Alidoust}}, \bibinfo {author} {\bibfnamefont {T.-R.}\ \bibnamefont {Chang}},
  \bibinfo {author} {\bibfnamefont {H.-T.}\ \bibnamefont {Jeng}}, \bibinfo
  {author} {\bibfnamefont {C.-Y.}\ \bibnamefont {Huang}}, \bibinfo {author}
  {\bibfnamefont {W.-F.}\ \bibnamefont {Tsai}}, \bibinfo {author}
  {\bibfnamefont {H.}~\bibnamefont {Lin}}, \bibinfo {author} {\bibfnamefont
  {P.~P.}\ \bibnamefont {Shibayev}}, \bibinfo {author} {\bibfnamefont {F.-C.}\
  \bibnamefont {Chou}}, \bibinfo {author} {\bibfnamefont {R.~J.}\ \bibnamefont
  {Cava}}, \ and\ \bibinfo {author} {\bibfnamefont {M.~Z.}\ \bibnamefont
  {Hasan}},\ }\href {\doibase 10.1126/science.1256742} {\bibfield  {journal}
  {\bibinfo  {journal} {Science}\ }\textbf {\bibinfo {volume} {347}},\ \bibinfo
  {pages} {294} (\bibinfo {year} {2015}{\natexlab{a}})},\ \Eprint
  {http://arxiv.org/abs/https://science.sciencemag.org/content/347/6219/294.full.pdf}
  {https://science.sciencemag.org/content/347/6219/294.full.pdf} \BibitemShut
  {NoStop}%
\bibitem [{\citenamefont {Lv}\ \emph {et~al.}(2015)\citenamefont {Lv},
  \citenamefont {Weng}, \citenamefont {Fu}, \citenamefont {Wang}, \citenamefont
  {Miao}, \citenamefont {Ma}, \citenamefont {Richard}, \citenamefont {Huang},
  \citenamefont {Zhao}, \citenamefont {Chen}, \citenamefont {Fang},
  \citenamefont {Dai}, \citenamefont {Qian},\ and\ \citenamefont
  {Ding}}]{lv15}%
  \BibitemOpen
  \bibfield  {author} {\bibinfo {author} {\bibfnamefont {B.~Q.}\ \bibnamefont
  {Lv}}, \bibinfo {author} {\bibfnamefont {H.~M.}\ \bibnamefont {Weng}},
  \bibinfo {author} {\bibfnamefont {B.~B.}\ \bibnamefont {Fu}}, \bibinfo
  {author} {\bibfnamefont {X.~P.}\ \bibnamefont {Wang}}, \bibinfo {author}
  {\bibfnamefont {H.}~\bibnamefont {Miao}}, \bibinfo {author} {\bibfnamefont
  {J.}~\bibnamefont {Ma}}, \bibinfo {author} {\bibfnamefont {P.}~\bibnamefont
  {Richard}}, \bibinfo {author} {\bibfnamefont {X.~C.}\ \bibnamefont {Huang}},
  \bibinfo {author} {\bibfnamefont {L.~X.}\ \bibnamefont {Zhao}}, \bibinfo
  {author} {\bibfnamefont {G.~F.}\ \bibnamefont {Chen}}, \bibinfo {author}
  {\bibfnamefont {Z.}~\bibnamefont {Fang}}, \bibinfo {author} {\bibfnamefont
  {X.}~\bibnamefont {Dai}}, \bibinfo {author} {\bibfnamefont {T.}~\bibnamefont
  {Qian}}, \ and\ \bibinfo {author} {\bibfnamefont {H.}~\bibnamefont {Ding}},\
  }\href {\doibase 10.1103/PhysRevX.5.031013} {\bibfield  {journal} {\bibinfo
  {journal} {Phys. Rev. X}\ }\textbf {\bibinfo {volume} {5}},\ \bibinfo {pages}
  {031013} (\bibinfo {year} {2015})}\BibitemShut {NoStop}%
\bibitem [{\citenamefont {Xu}\ \emph {et~al.}(2015{\natexlab{b}})\citenamefont
  {Xu}, \citenamefont {Belopolski}, \citenamefont {Alidoust}, \citenamefont
  {Neupane}, \citenamefont {Bian}, \citenamefont {Zhang}, \citenamefont
  {Sankar}, \citenamefont {Chang}, \citenamefont {Yuan}, \citenamefont {Lee},
  \citenamefont {Huang}, \citenamefont {Zheng}, \citenamefont {Ma},
  \citenamefont {Sanchez}, \citenamefont {Wang}, \citenamefont {Bansil},
  \citenamefont {Chou}, \citenamefont {Shibayev}, \citenamefont {Lin},
  \citenamefont {Jia},\ and\ \citenamefont {Hasan}}]{xu15_2}%
  \BibitemOpen
  \bibfield  {author} {\bibinfo {author} {\bibfnamefont {S.-Y.}\ \bibnamefont
  {Xu}}, \bibinfo {author} {\bibfnamefont {I.}~\bibnamefont {Belopolski}},
  \bibinfo {author} {\bibfnamefont {N.}~\bibnamefont {Alidoust}}, \bibinfo
  {author} {\bibfnamefont {M.}~\bibnamefont {Neupane}}, \bibinfo {author}
  {\bibfnamefont {G.}~\bibnamefont {Bian}}, \bibinfo {author} {\bibfnamefont
  {C.}~\bibnamefont {Zhang}}, \bibinfo {author} {\bibfnamefont
  {R.}~\bibnamefont {Sankar}}, \bibinfo {author} {\bibfnamefont
  {G.}~\bibnamefont {Chang}}, \bibinfo {author} {\bibfnamefont
  {Z.}~\bibnamefont {Yuan}}, \bibinfo {author} {\bibfnamefont {C.-C.}\
  \bibnamefont {Lee}}, \bibinfo {author} {\bibfnamefont {S.-M.}\ \bibnamefont
  {Huang}}, \bibinfo {author} {\bibfnamefont {H.}~\bibnamefont {Zheng}},
  \bibinfo {author} {\bibfnamefont {J.}~\bibnamefont {Ma}}, \bibinfo {author}
  {\bibfnamefont {D.~S.}\ \bibnamefont {Sanchez}}, \bibinfo {author}
  {\bibfnamefont {B.}~\bibnamefont {Wang}}, \bibinfo {author} {\bibfnamefont
  {A.}~\bibnamefont {Bansil}}, \bibinfo {author} {\bibfnamefont
  {F.}~\bibnamefont {Chou}}, \bibinfo {author} {\bibfnamefont {P.~P.}\
  \bibnamefont {Shibayev}}, \bibinfo {author} {\bibfnamefont {H.}~\bibnamefont
  {Lin}}, \bibinfo {author} {\bibfnamefont {S.}~\bibnamefont {Jia}}, \ and\
  \bibinfo {author} {\bibfnamefont {M.~Z.}\ \bibnamefont {Hasan}},\ }\href
  {\doibase 10.1126/science.aaa9297} {\bibfield  {journal} {\bibinfo  {journal}
  {Science}\ }\textbf {\bibinfo {volume} {349}},\ \bibinfo {pages} {613}
  (\bibinfo {year} {2015}{\natexlab{b}})},\ \Eprint
  {http://arxiv.org/abs/https://science.sciencemag.org/content/349/6248/613.full.pdf}
  {https://science.sciencemag.org/content/349/6248/613.full.pdf} \BibitemShut
  {NoStop}%
\bibitem [{\citenamefont {Belopolski}\ \emph {et~al.}(2016)\citenamefont
  {Belopolski}, \citenamefont {Xu}, \citenamefont {Sanchez}, \citenamefont
  {Chang}, \citenamefont {Guo}, \citenamefont {Neupane}, \citenamefont {Zheng},
  \citenamefont {Lee}, \citenamefont {Huang}, \citenamefont {Bian},
  \citenamefont {Alidoust}, \citenamefont {Chang}, \citenamefont {Wang},
  \citenamefont {Zhang}, \citenamefont {Bansil}, \citenamefont {Jeng},
  \citenamefont {Lin}, \citenamefont {Jia},\ and\ \citenamefont
  {Hasan}}]{belopolski16}%
  \BibitemOpen
  \bibfield  {author} {\bibinfo {author} {\bibfnamefont {I.}~\bibnamefont
  {Belopolski}}, \bibinfo {author} {\bibfnamefont {S.-Y.}\ \bibnamefont {Xu}},
  \bibinfo {author} {\bibfnamefont {D.~S.}\ \bibnamefont {Sanchez}}, \bibinfo
  {author} {\bibfnamefont {G.}~\bibnamefont {Chang}}, \bibinfo {author}
  {\bibfnamefont {C.}~\bibnamefont {Guo}}, \bibinfo {author} {\bibfnamefont
  {M.}~\bibnamefont {Neupane}}, \bibinfo {author} {\bibfnamefont
  {H.}~\bibnamefont {Zheng}}, \bibinfo {author} {\bibfnamefont {C.-C.}\
  \bibnamefont {Lee}}, \bibinfo {author} {\bibfnamefont {S.-M.}\ \bibnamefont
  {Huang}}, \bibinfo {author} {\bibfnamefont {G.}~\bibnamefont {Bian}},
  \bibinfo {author} {\bibfnamefont {N.}~\bibnamefont {Alidoust}}, \bibinfo
  {author} {\bibfnamefont {T.-R.}\ \bibnamefont {Chang}}, \bibinfo {author}
  {\bibfnamefont {B.}~\bibnamefont {Wang}}, \bibinfo {author} {\bibfnamefont
  {X.}~\bibnamefont {Zhang}}, \bibinfo {author} {\bibfnamefont
  {A.}~\bibnamefont {Bansil}}, \bibinfo {author} {\bibfnamefont {H.-T.}\
  \bibnamefont {Jeng}}, \bibinfo {author} {\bibfnamefont {H.}~\bibnamefont
  {Lin}}, \bibinfo {author} {\bibfnamefont {S.}~\bibnamefont {Jia}}, \ and\
  \bibinfo {author} {\bibfnamefont {M.~Z.}\ \bibnamefont {Hasan}},\ }\href
  {\doibase 10.1103/PhysRevLett.116.066802} {\bibfield  {journal} {\bibinfo
  {journal} {Phys. Rev. Lett.}\ }\textbf {\bibinfo {volume} {116}},\ \bibinfo
  {pages} {066802} (\bibinfo {year} {2016})}\BibitemShut {NoStop}%
\bibitem [{\citenamefont {Inoue}\ \emph {et~al.}(2016)\citenamefont {Inoue},
  \citenamefont {Gyenis}, \citenamefont {Wang}, \citenamefont {Li},
  \citenamefont {Oh}, \citenamefont {Jiang}, \citenamefont {Ni}, \citenamefont
  {Bernevig},\ and\ \citenamefont {Yazdani}}]{inoue16}%
  \BibitemOpen
  \bibfield  {author} {\bibinfo {author} {\bibfnamefont {H.}~\bibnamefont
  {Inoue}}, \bibinfo {author} {\bibfnamefont {A.}~\bibnamefont {Gyenis}},
  \bibinfo {author} {\bibfnamefont {Z.}~\bibnamefont {Wang}}, \bibinfo {author}
  {\bibfnamefont {J.}~\bibnamefont {Li}}, \bibinfo {author} {\bibfnamefont
  {S.~W.}\ \bibnamefont {Oh}}, \bibinfo {author} {\bibfnamefont
  {S.}~\bibnamefont {Jiang}}, \bibinfo {author} {\bibfnamefont
  {N.}~\bibnamefont {Ni}}, \bibinfo {author} {\bibfnamefont {B.~A.}\
  \bibnamefont {Bernevig}}, \ and\ \bibinfo {author} {\bibfnamefont
  {A.}~\bibnamefont {Yazdani}},\ }\href {\doibase 10.1126/science.aad8766}
  {\bibfield  {journal} {\bibinfo  {journal} {Science}\ }\textbf {\bibinfo
  {volume} {351}},\ \bibinfo {pages} {1184} (\bibinfo {year} {2016})},\ \Eprint
  {http://arxiv.org/abs/https://science.sciencemag.org/content/351/6278/1184.full.pdf}
  {https://science.sciencemag.org/content/351/6278/1184.full.pdf} \BibitemShut
  {NoStop}%
\bibitem [{\citenamefont {Potter}\ \emph {et~al.}(2014)\citenamefont {Potter},
  \citenamefont {Kimchi},\ and\ \citenamefont {Vishwanath}}]{potter2014}%
  \BibitemOpen
  \bibfield  {author} {\bibinfo {author} {\bibfnamefont {A.~C.}\ \bibnamefont
  {Potter}}, \bibinfo {author} {\bibfnamefont {I.}~\bibnamefont {Kimchi}}, \
  and\ \bibinfo {author} {\bibfnamefont {A.}~\bibnamefont {Vishwanath}},\
  }\href@noop {} {\bibfield  {journal} {\bibinfo  {journal} {Nature
  communications}\ }\textbf {\bibinfo {volume} {5}},\ \bibinfo {pages} {1}
  (\bibinfo {year} {2014})}\BibitemShut {NoStop}%
\bibitem [{\citenamefont {Moll}\ \emph {et~al.}(2016)\citenamefont {Moll},
  \citenamefont {Nair}, \citenamefont {Helm}, \citenamefont {Potter},
  \citenamefont {Kimchi}, \citenamefont {Vishwanath},\ and\ \citenamefont
  {Analytis}}]{moll16}%
  \BibitemOpen
  \bibfield  {author} {\bibinfo {author} {\bibfnamefont {P.~J.~W.}\
  \bibnamefont {Moll}}, \bibinfo {author} {\bibfnamefont {N.~L.}\ \bibnamefont
  {Nair}}, \bibinfo {author} {\bibfnamefont {T.}~\bibnamefont {Helm}}, \bibinfo
  {author} {\bibfnamefont {A.~C.}\ \bibnamefont {Potter}}, \bibinfo {author}
  {\bibfnamefont {I.}~\bibnamefont {Kimchi}}, \bibinfo {author} {\bibfnamefont
  {A.}~\bibnamefont {Vishwanath}}, \ and\ \bibinfo {author} {\bibfnamefont
  {J.~G.}\ \bibnamefont {Analytis}},\ }\href {\doibase 10.1038/nature18276}
  {\bibfield  {journal} {\bibinfo  {journal} {Nature}\ }\textbf {\bibinfo
  {volume} {535}},\ \bibinfo {pages} {266} (\bibinfo {year}
  {2016})}\BibitemShut {NoStop}%
\bibitem [{\citenamefont {Jia}\ \emph {et~al.}(2016)\citenamefont {Jia},
  \citenamefont {Xu},\ and\ \citenamefont {Hasan}}]{jia16}%
  \BibitemOpen
  \bibfield  {author} {\bibinfo {author} {\bibfnamefont {S.}~\bibnamefont
  {Jia}}, \bibinfo {author} {\bibfnamefont {S.-Y.}\ \bibnamefont {Xu}}, \ and\
  \bibinfo {author} {\bibfnamefont {M.~Z.}\ \bibnamefont {Hasan}},\ }\href
  {\doibase 10.1038/nmat4787} {\bibfield  {journal} {\bibinfo  {journal}
  {Nature Materials}\ }\textbf {\bibinfo {volume} {15}},\ \bibinfo {pages}
  {1140} (\bibinfo {year} {2016})}\BibitemShut {NoStop}%
\bibitem [{\citenamefont {Shi}\ and\ \citenamefont {Song}(2017)}]{shi2017}%
  \BibitemOpen
  \bibfield  {author} {\bibinfo {author} {\bibfnamefont {L.-k.}\ \bibnamefont
  {Shi}}\ and\ \bibinfo {author} {\bibfnamefont {J.~C.~W.}\ \bibnamefont
  {Song}},\ }\href {\doibase 10.1103/PhysRevB.96.081410} {\bibfield  {journal}
  {\bibinfo  {journal} {Phys. Rev. B}\ }\textbf {\bibinfo {volume} {96}},\
  \bibinfo {pages} {081410} (\bibinfo {year} {2017})}\BibitemShut {NoStop}%
\bibitem [{\citenamefont {Song}\ and\ \citenamefont {Rudner}(2017)}]{song2017}%
  \BibitemOpen
  \bibfield  {author} {\bibinfo {author} {\bibfnamefont {J.~C.~W.}\
  \bibnamefont {Song}}\ and\ \bibinfo {author} {\bibfnamefont {M.~S.}\
  \bibnamefont {Rudner}},\ }\href {\doibase 10.1103/PhysRevB.96.205443}
  {\bibfield  {journal} {\bibinfo  {journal} {Phys. Rev. B}\ }\textbf {\bibinfo
  {volume} {96}},\ \bibinfo {pages} {205443} (\bibinfo {year}
  {2017})}\BibitemShut {NoStop}%
\bibitem [{\citenamefont {Mukherjee}\ \emph {et~al.}(2019)\citenamefont
  {Mukherjee}, \citenamefont {Carpentier},\ and\ \citenamefont
  {Goerbig}}]{Mukherjee2019}%
  \BibitemOpen
  \bibfield  {author} {\bibinfo {author} {\bibfnamefont {D.~K.}\ \bibnamefont
  {Mukherjee}}, \bibinfo {author} {\bibfnamefont {D.}~\bibnamefont
  {Carpentier}}, \ and\ \bibinfo {author} {\bibfnamefont {M.~O.}\ \bibnamefont
  {Goerbig}},\ }\href {\doibase 10.1103/PhysRevB.100.195412} {\bibfield
  {journal} {\bibinfo  {journal} {Phys. Rev. B}\ }\textbf {\bibinfo {volume}
  {100}},\ \bibinfo {pages} {195412} (\bibinfo {year} {2019})}\BibitemShut
  {NoStop}%
\bibitem [{\citenamefont {Ghosh}\ and\ \citenamefont {Timm}(2020)}]{ghosh2020}%
  \BibitemOpen
  \bibfield  {author} {\bibinfo {author} {\bibfnamefont {S.}~\bibnamefont
  {Ghosh}}\ and\ \bibinfo {author} {\bibfnamefont {C.}~\bibnamefont {Timm}},\
  }\href {\doibase 10.1103/PhysRevB.101.165402} {\bibfield  {journal} {\bibinfo
   {journal} {Phys. Rev. B}\ }\textbf {\bibinfo {volume} {101}},\ \bibinfo
  {pages} {165402} (\bibinfo {year} {2020})}\BibitemShut {NoStop}%
\bibitem [{\citenamefont {Wawrzik}\ \emph {et~al.}(2020)\citenamefont
  {Wawrzik}, \citenamefont {You}, \citenamefont {Facio}, \citenamefont {van~den
  Brink},\ and\ \citenamefont {Sodemann}}]{wawrzik20}%
  \BibitemOpen
  \bibfield  {author} {\bibinfo {author} {\bibfnamefont {D.}~\bibnamefont
  {Wawrzik}}, \bibinfo {author} {\bibfnamefont {J.-S.}\ \bibnamefont {You}},
  \bibinfo {author} {\bibfnamefont {J.~I.}\ \bibnamefont {Facio}}, \bibinfo
  {author} {\bibfnamefont {J.}~\bibnamefont {van~den Brink}}, \ and\ \bibinfo
  {author} {\bibfnamefont {I.}~\bibnamefont {Sodemann}},\ }\href@noop {}
  {\enquote {\bibinfo {title} {The infinite berry curvature of weyl fermi
  arcs},}\ } (\bibinfo {year} {2020}),\ \Eprint
  {http://arxiv.org/abs/2010.10537} {arXiv:2010.10537 [cond-mat.mes-hall]}
  \BibitemShut {NoStop}%
\bibitem [{\citenamefont {Chang}\ \emph {et~al.}(2017)\citenamefont {Chang},
  \citenamefont {Xu}, \citenamefont {Wieder}, \citenamefont {Sanchez},
  \citenamefont {Huang}, \citenamefont {Belopolski}, \citenamefont {Chang},
  \citenamefont {Zhang}, \citenamefont {Bansil}, \citenamefont {Lin},\ and\
  \citenamefont {Hasan}}]{chang17}%
  \BibitemOpen
  \bibfield  {author} {\bibinfo {author} {\bibfnamefont {G.}~\bibnamefont
  {Chang}}, \bibinfo {author} {\bibfnamefont {S.-Y.}\ \bibnamefont {Xu}},
  \bibinfo {author} {\bibfnamefont {B.~J.}\ \bibnamefont {Wieder}}, \bibinfo
  {author} {\bibfnamefont {D.~S.}\ \bibnamefont {Sanchez}}, \bibinfo {author}
  {\bibfnamefont {S.-M.}\ \bibnamefont {Huang}}, \bibinfo {author}
  {\bibfnamefont {I.}~\bibnamefont {Belopolski}}, \bibinfo {author}
  {\bibfnamefont {T.-R.}\ \bibnamefont {Chang}}, \bibinfo {author}
  {\bibfnamefont {S.}~\bibnamefont {Zhang}}, \bibinfo {author} {\bibfnamefont
  {A.}~\bibnamefont {Bansil}}, \bibinfo {author} {\bibfnamefont
  {H.}~\bibnamefont {Lin}}, \ and\ \bibinfo {author} {\bibfnamefont {M.~Z.}\
  \bibnamefont {Hasan}},\ }\href {\doibase 10.1103/PhysRevLett.119.206401}
  {\bibfield  {journal} {\bibinfo  {journal} {Phys. Rev. Lett.}\ }\textbf
  {\bibinfo {volume} {119}},\ \bibinfo {pages} {206401} (\bibinfo {year}
  {2017})}\BibitemShut {NoStop}%
\bibitem [{\citenamefont {Tang}\ \emph {et~al.}(2017)\citenamefont {Tang},
  \citenamefont {Zhou},\ and\ \citenamefont {Zhang}}]{tang17}%
  \BibitemOpen
  \bibfield  {author} {\bibinfo {author} {\bibfnamefont {P.}~\bibnamefont
  {Tang}}, \bibinfo {author} {\bibfnamefont {Q.}~\bibnamefont {Zhou}}, \ and\
  \bibinfo {author} {\bibfnamefont {S.-C.}\ \bibnamefont {Zhang}},\ }\href
  {\doibase 10.1103/PhysRevLett.119.206402} {\bibfield  {journal} {\bibinfo
  {journal} {Phys. Rev. Lett.}\ }\textbf {\bibinfo {volume} {119}},\ \bibinfo
  {pages} {206402} (\bibinfo {year} {2017})}\BibitemShut {NoStop}%
\bibitem [{\citenamefont {Chang}\ \emph {et~al.}(2018)\citenamefont {Chang},
  \citenamefont {Wieder}, \citenamefont {Schindler}, \citenamefont {Sanchez},
  \citenamefont {Belopolski}, \citenamefont {Huang}, \citenamefont {Singh},
  \citenamefont {Wu}, \citenamefont {Chang}, \citenamefont {Neupert},
  \citenamefont {Xu}, \citenamefont {Lin},\ and\ \citenamefont
  {Hasan}}]{chang18}%
  \BibitemOpen
  \bibfield  {author} {\bibinfo {author} {\bibfnamefont {G.}~\bibnamefont
  {Chang}}, \bibinfo {author} {\bibfnamefont {B.~J.}\ \bibnamefont {Wieder}},
  \bibinfo {author} {\bibfnamefont {F.}~\bibnamefont {Schindler}}, \bibinfo
  {author} {\bibfnamefont {D.~S.}\ \bibnamefont {Sanchez}}, \bibinfo {author}
  {\bibfnamefont {I.}~\bibnamefont {Belopolski}}, \bibinfo {author}
  {\bibfnamefont {S.-M.}\ \bibnamefont {Huang}}, \bibinfo {author}
  {\bibfnamefont {B.}~\bibnamefont {Singh}}, \bibinfo {author} {\bibfnamefont
  {D.}~\bibnamefont {Wu}}, \bibinfo {author} {\bibfnamefont {T.-R.}\
  \bibnamefont {Chang}}, \bibinfo {author} {\bibfnamefont {T.}~\bibnamefont
  {Neupert}}, \bibinfo {author} {\bibfnamefont {S.-Y.}\ \bibnamefont {Xu}},
  \bibinfo {author} {\bibfnamefont {H.}~\bibnamefont {Lin}}, \ and\ \bibinfo
  {author} {\bibfnamefont {M.~Z.}\ \bibnamefont {Hasan}},\ }\href {\doibase
  10.1038/s41563-018-0169-3} {\bibfield  {journal} {\bibinfo  {journal} {Nature
  Materials}\ }\textbf {\bibinfo {volume} {17}},\ \bibinfo {pages} {978}
  (\bibinfo {year} {2018})}\BibitemShut {NoStop}%
\bibitem [{\citenamefont {Flicker}\ \emph {et~al.}(2018)\citenamefont
  {Flicker}, \citenamefont {de~Juan}, \citenamefont {Bradlyn}, \citenamefont
  {Morimoto}, \citenamefont {Vergniory},\ and\ \citenamefont
  {Grushin}}]{flicker18}%
  \BibitemOpen
  \bibfield  {author} {\bibinfo {author} {\bibfnamefont {F.}~\bibnamefont
  {Flicker}}, \bibinfo {author} {\bibfnamefont {F.}~\bibnamefont {de~Juan}},
  \bibinfo {author} {\bibfnamefont {B.}~\bibnamefont {Bradlyn}}, \bibinfo
  {author} {\bibfnamefont {T.}~\bibnamefont {Morimoto}}, \bibinfo {author}
  {\bibfnamefont {M.~G.}\ \bibnamefont {Vergniory}}, \ and\ \bibinfo {author}
  {\bibfnamefont {A.~G.}\ \bibnamefont {Grushin}},\ }\href {\doibase
  10.1103/PhysRevB.98.155145} {\bibfield  {journal} {\bibinfo  {journal} {Phys.
  Rev. B}\ }\textbf {\bibinfo {volume} {98}},\ \bibinfo {pages} {155145}
  (\bibinfo {year} {2018})}\BibitemShut {NoStop}%
\bibitem [{\citenamefont {Le}\ \emph {et~al.}(2020)\citenamefont {Le},
  \citenamefont {Zhang}, \citenamefont {Felser},\ and\ \citenamefont
  {Sun}}]{le20}%
  \BibitemOpen
  \bibfield  {author} {\bibinfo {author} {\bibfnamefont {C.}~\bibnamefont
  {Le}}, \bibinfo {author} {\bibfnamefont {Y.}~\bibnamefont {Zhang}}, \bibinfo
  {author} {\bibfnamefont {C.}~\bibnamefont {Felser}}, \ and\ \bibinfo {author}
  {\bibfnamefont {Y.}~\bibnamefont {Sun}},\ }\href {\doibase
  10.1103/PhysRevB.102.121111} {\bibfield  {journal} {\bibinfo  {journal}
  {Phys. Rev. B}\ }\textbf {\bibinfo {volume} {102}},\ \bibinfo {pages}
  {121111} (\bibinfo {year} {2020})}\BibitemShut {NoStop}%
\bibitem [{\citenamefont {Ni}\ \emph {et~al.}(2020)\citenamefont {Ni},
  \citenamefont {Xu}, \citenamefont {S{\'{a}}nchez-Mart{\'{i}}nez},
  \citenamefont {Zhang}, \citenamefont {Manna}, \citenamefont {Bernhard},
  \citenamefont {Venderbos}, \citenamefont {de~Juan}, \citenamefont {Felser},
  \citenamefont {Grushin},\ and\ \citenamefont {Wu}}]{ni20}%
  \BibitemOpen
  \bibfield  {author} {\bibinfo {author} {\bibfnamefont {Z.}~\bibnamefont
  {Ni}}, \bibinfo {author} {\bibfnamefont {B.}~\bibnamefont {Xu}}, \bibinfo
  {author} {\bibfnamefont {M.-{\'{A}}.}\ \bibnamefont
  {S{\'{a}}nchez-Mart{\'{i}}nez}}, \bibinfo {author} {\bibfnamefont
  {Y.}~\bibnamefont {Zhang}}, \bibinfo {author} {\bibfnamefont
  {K.}~\bibnamefont {Manna}}, \bibinfo {author} {\bibfnamefont
  {C.}~\bibnamefont {Bernhard}}, \bibinfo {author} {\bibfnamefont {J.~W.~F.}\
  \bibnamefont {Venderbos}}, \bibinfo {author} {\bibfnamefont {F.}~\bibnamefont
  {de~Juan}}, \bibinfo {author} {\bibfnamefont {C.}~\bibnamefont {Felser}},
  \bibinfo {author} {\bibfnamefont {A.~G.}\ \bibnamefont {Grushin}}, \ and\
  \bibinfo {author} {\bibfnamefont {L.}~\bibnamefont {Wu}},\ }\href {\doibase
  10.1038/s41535-020-00298-y} {\bibfield  {journal} {\bibinfo  {journal} {npj
  Quantum Materials}\ }\textbf {\bibinfo {volume} {5}},\ \bibinfo {pages} {96}
  (\bibinfo {year} {2020})}\BibitemShut {NoStop}%
\bibitem [{\citenamefont {Li}\ \emph {et~al.}(2019)\citenamefont {Li},
  \citenamefont {Xu}, \citenamefont {Rao}, \citenamefont {Zhou}, \citenamefont
  {Wang}, \citenamefont {Zhou}, \citenamefont {Tian}, \citenamefont {Gao},
  \citenamefont {Li}, \citenamefont {Huang}, \citenamefont {Lei}, \citenamefont
  {Weng}, \citenamefont {Sun}, \citenamefont {Xia}, \citenamefont {Qian},\ and\
  \citenamefont {Ding}}]{li19}%
  \BibitemOpen
  \bibfield  {author} {\bibinfo {author} {\bibfnamefont {H.}~\bibnamefont
  {Li}}, \bibinfo {author} {\bibfnamefont {S.}~\bibnamefont {Xu}}, \bibinfo
  {author} {\bibfnamefont {Z.-C.}\ \bibnamefont {Rao}}, \bibinfo {author}
  {\bibfnamefont {L.-Q.}\ \bibnamefont {Zhou}}, \bibinfo {author}
  {\bibfnamefont {Z.-J.}\ \bibnamefont {Wang}}, \bibinfo {author}
  {\bibfnamefont {S.-M.}\ \bibnamefont {Zhou}}, \bibinfo {author}
  {\bibfnamefont {S.-J.}\ \bibnamefont {Tian}}, \bibinfo {author}
  {\bibfnamefont {S.-Y.}\ \bibnamefont {Gao}}, \bibinfo {author} {\bibfnamefont
  {J.-J.}\ \bibnamefont {Li}}, \bibinfo {author} {\bibfnamefont {Y.-B.}\
  \bibnamefont {Huang}}, \bibinfo {author} {\bibfnamefont {H.-C.}\ \bibnamefont
  {Lei}}, \bibinfo {author} {\bibfnamefont {H.-M.}\ \bibnamefont {Weng}},
  \bibinfo {author} {\bibfnamefont {Y.-J.}\ \bibnamefont {Sun}}, \bibinfo
  {author} {\bibfnamefont {T.-L.}\ \bibnamefont {Xia}}, \bibinfo {author}
  {\bibfnamefont {T.}~\bibnamefont {Qian}}, \ and\ \bibinfo {author}
  {\bibfnamefont {H.}~\bibnamefont {Ding}},\ }\href {\doibase
  10.1038/s41467-019-13435-4} {\bibfield  {journal} {\bibinfo  {journal}
  {Nature Communications}\ }\textbf {\bibinfo {volume} {10}},\ \bibinfo {pages}
  {5505} (\bibinfo {year} {2019})}\BibitemShut {NoStop}%
\bibitem [{\citenamefont {Cochran}\ \emph {et~al.}(2020)\citenamefont
  {Cochran}, \citenamefont {Chang}, \citenamefont {Belopolski}, \citenamefont
  {Manna}, \citenamefont {Sanchez}, \citenamefont {Chéng}, \citenamefont
  {Yin}, \citenamefont {Borrmann}, \citenamefont {Denlinger}, \citenamefont
  {Felser}, \citenamefont {Lin},\ and\ \citenamefont {Hasan}}]{cochran20}%
  \BibitemOpen
  \bibfield  {author} {\bibinfo {author} {\bibfnamefont {T.~A.}\ \bibnamefont
  {Cochran}}, \bibinfo {author} {\bibfnamefont {G.}~\bibnamefont {Chang}},
  \bibinfo {author} {\bibfnamefont {I.}~\bibnamefont {Belopolski}}, \bibinfo
  {author} {\bibfnamefont {K.}~\bibnamefont {Manna}}, \bibinfo {author}
  {\bibfnamefont {D.~S.}\ \bibnamefont {Sanchez}}, \bibinfo {author}
  {\bibfnamefont {Z.}~\bibnamefont {Chéng}}, \bibinfo {author} {\bibfnamefont
  {J.-X.}\ \bibnamefont {Yin}}, \bibinfo {author} {\bibfnamefont
  {H.}~\bibnamefont {Borrmann}}, \bibinfo {author} {\bibfnamefont
  {J.}~\bibnamefont {Denlinger}}, \bibinfo {author} {\bibfnamefont
  {C.}~\bibnamefont {Felser}}, \bibinfo {author} {\bibfnamefont
  {H.}~\bibnamefont {Lin}}, \ and\ \bibinfo {author} {\bibfnamefont {M.~Z.}\
  \bibnamefont {Hasan}},\ }\href@noop {} {\bibfield  {journal} {\bibinfo
  {journal} {arXiv}\ } (\bibinfo {year} {2020})},\ \Eprint
  {http://arxiv.org/abs/2004.11365} {arXiv:2004.11365 [cond-mat.mtrl-sci]}
  \BibitemShut {NoStop}%
\bibitem [{\citenamefont {Sanchez}\ \emph {et~al.}(2019)\citenamefont
  {Sanchez}, \citenamefont {Belopolski}, \citenamefont {Cochran}, \citenamefont
  {Xu}, \citenamefont {Yin}, \citenamefont {Chang}, \citenamefont {Xie},
  \citenamefont {Manna}, \citenamefont {S{\"{u}}{\ss}}, \citenamefont {Huang},
  \citenamefont {Alidoust}, \citenamefont {Multer}, \citenamefont {Zhang},
  \citenamefont {Shumiya}, \citenamefont {Wang}, \citenamefont {Wang},
  \citenamefont {Chang}, \citenamefont {Felser}, \citenamefont {Xu},
  \citenamefont {Jia}, \citenamefont {Lin},\ and\ \citenamefont
  {Hasan}}]{sanchez19}%
  \BibitemOpen
  \bibfield  {author} {\bibinfo {author} {\bibfnamefont {D.~S.}\ \bibnamefont
  {Sanchez}}, \bibinfo {author} {\bibfnamefont {I.}~\bibnamefont {Belopolski}},
  \bibinfo {author} {\bibfnamefont {T.~A.}\ \bibnamefont {Cochran}}, \bibinfo
  {author} {\bibfnamefont {X.}~\bibnamefont {Xu}}, \bibinfo {author}
  {\bibfnamefont {J.~X.}\ \bibnamefont {Yin}}, \bibinfo {author} {\bibfnamefont
  {G.}~\bibnamefont {Chang}}, \bibinfo {author} {\bibfnamefont
  {W.}~\bibnamefont {Xie}}, \bibinfo {author} {\bibfnamefont {K.}~\bibnamefont
  {Manna}}, \bibinfo {author} {\bibfnamefont {V.}~\bibnamefont
  {S{\"{u}}{\ss}}}, \bibinfo {author} {\bibfnamefont {C.~Y.}\ \bibnamefont
  {Huang}}, \bibinfo {author} {\bibfnamefont {N.}~\bibnamefont {Alidoust}},
  \bibinfo {author} {\bibfnamefont {D.}~\bibnamefont {Multer}}, \bibinfo
  {author} {\bibfnamefont {S.~S.}\ \bibnamefont {Zhang}}, \bibinfo {author}
  {\bibfnamefont {N.}~\bibnamefont {Shumiya}}, \bibinfo {author} {\bibfnamefont
  {X.}~\bibnamefont {Wang}}, \bibinfo {author} {\bibfnamefont {G.~Q.}\
  \bibnamefont {Wang}}, \bibinfo {author} {\bibfnamefont {T.~R.}\ \bibnamefont
  {Chang}}, \bibinfo {author} {\bibfnamefont {C.}~\bibnamefont {Felser}},
  \bibinfo {author} {\bibfnamefont {S.~Y.}\ \bibnamefont {Xu}}, \bibinfo
  {author} {\bibfnamefont {S.}~\bibnamefont {Jia}}, \bibinfo {author}
  {\bibfnamefont {H.}~\bibnamefont {Lin}}, \ and\ \bibinfo {author}
  {\bibfnamefont {M.~Z.}\ \bibnamefont {Hasan}},\ }\href {\doibase
  10.1038/s41586-019-1037-2} {\bibfield  {journal} {\bibinfo  {journal}
  {Nature}\ }\textbf {\bibinfo {volume} {567}},\ \bibinfo {pages} {500}
  (\bibinfo {year} {2019})}\BibitemShut {NoStop}%
\bibitem [{\citenamefont {de~Juan}\ \emph {et~al.}(2017)\citenamefont
  {de~Juan}, \citenamefont {Grushin}, \citenamefont {Morimoto},\ and\
  \citenamefont {Moore}}]{dejuan17}%
  \BibitemOpen
  \bibfield  {author} {\bibinfo {author} {\bibfnamefont {F.}~\bibnamefont
  {de~Juan}}, \bibinfo {author} {\bibfnamefont {A.~G.}\ \bibnamefont
  {Grushin}}, \bibinfo {author} {\bibfnamefont {T.}~\bibnamefont {Morimoto}}, \
  and\ \bibinfo {author} {\bibfnamefont {J.~E.}\ \bibnamefont {Moore}},\ }\href
  {https://doi.org/10.1038/ncomms15995 http://10.0.4.14/ncomms15995
  https://www.nature.com/articles/ncomms15995{\#}supplementary-information}
  {\bibfield  {journal} {\bibinfo  {journal} {Nature Communications}\ }\textbf
  {\bibinfo {volume} {8}},\ \bibinfo {pages} {15995} (\bibinfo {year}
  {2017})}\BibitemShut {NoStop}%
\bibitem [{\citenamefont {S\'anchez-Mart\'{\i}nez}\ \emph
  {et~al.}(2019)\citenamefont {S\'anchez-Mart\'{\i}nez}, \citenamefont
  {de~Juan},\ and\ \citenamefont {Grushin}}]{martinez19}%
  \BibitemOpen
  \bibfield  {author} {\bibinfo {author} {\bibfnamefont {M.-A.}\ \bibnamefont
  {S\'anchez-Mart\'{\i}nez}}, \bibinfo {author} {\bibfnamefont
  {F.}~\bibnamefont {de~Juan}}, \ and\ \bibinfo {author} {\bibfnamefont
  {A.~G.}\ \bibnamefont {Grushin}},\ }\href {\doibase
  10.1103/PhysRevB.99.155145} {\bibfield  {journal} {\bibinfo  {journal} {Phys.
  Rev. B}\ }\textbf {\bibinfo {volume} {99}},\ \bibinfo {pages} {155145}
  (\bibinfo {year} {2019})}\BibitemShut {NoStop}%
\bibitem [{\citenamefont {Bradlyn}\ \emph {et~al.}(2016)\citenamefont
  {Bradlyn}, \citenamefont {Cano}, \citenamefont {Wang}, \citenamefont
  {Vergniory}, \citenamefont {Felser}, \citenamefont {Cava},\ and\
  \citenamefont {Bernevig}}]{bradlyn16}%
  \BibitemOpen
  \bibfield  {author} {\bibinfo {author} {\bibfnamefont {B.}~\bibnamefont
  {Bradlyn}}, \bibinfo {author} {\bibfnamefont {J.}~\bibnamefont {Cano}},
  \bibinfo {author} {\bibfnamefont {Z.}~\bibnamefont {Wang}}, \bibinfo {author}
  {\bibfnamefont {M.~G.}\ \bibnamefont {Vergniory}}, \bibinfo {author}
  {\bibfnamefont {C.}~\bibnamefont {Felser}}, \bibinfo {author} {\bibfnamefont
  {R.~J.}\ \bibnamefont {Cava}}, \ and\ \bibinfo {author} {\bibfnamefont
  {B.~A.}\ \bibnamefont {Bernevig}},\ }\href {\doibase 10.1126/science.aaf5037}
  {\bibfield  {journal} {\bibinfo  {journal} {Science}\ }\textbf {\bibinfo
  {volume} {353}} (\bibinfo {year} {2016}),\ 10.1126/science.aaf5037},\ \Eprint
  {http://arxiv.org/abs/https://science.sciencemag.org/content/353/6299/aaf5037.full.pdf}
  {https://science.sciencemag.org/content/353/6299/aaf5037.full.pdf}
  \BibitemShut {NoStop}%
\bibitem [{\citenamefont {Rees}\ \emph {et~al.}(2020)\citenamefont {Rees},
  \citenamefont {Manna}, \citenamefont {Lu}, \citenamefont {Morimoto},
  \citenamefont {Borrmann}, \citenamefont {Felser}, \citenamefont {Moore},
  \citenamefont {Torchinsky},\ and\ \citenamefont {Orenstein}}]{rees20}%
  \BibitemOpen
  \bibfield  {author} {\bibinfo {author} {\bibfnamefont {D.}~\bibnamefont
  {Rees}}, \bibinfo {author} {\bibfnamefont {K.}~\bibnamefont {Manna}},
  \bibinfo {author} {\bibfnamefont {B.}~\bibnamefont {Lu}}, \bibinfo {author}
  {\bibfnamefont {T.}~\bibnamefont {Morimoto}}, \bibinfo {author}
  {\bibfnamefont {H.}~\bibnamefont {Borrmann}}, \bibinfo {author}
  {\bibfnamefont {C.}~\bibnamefont {Felser}}, \bibinfo {author} {\bibfnamefont
  {J.~E.}\ \bibnamefont {Moore}}, \bibinfo {author} {\bibfnamefont {D.~H.}\
  \bibnamefont {Torchinsky}}, \ and\ \bibinfo {author} {\bibfnamefont
  {J.}~\bibnamefont {Orenstein}},\ }\href {\doibase 10.1126/sciadv.aba0509}
  {\bibfield  {journal} {\bibinfo  {journal} {Science Advances}\ }\textbf
  {\bibinfo {volume} {6}} (\bibinfo {year} {2020}),\ 10.1126/sciadv.aba0509},\
  \Eprint
  {http://arxiv.org/abs/https://advances.sciencemag.org/content/6/29/eaba0509.full.pdf}
  {https://advances.sciencemag.org/content/6/29/eaba0509.full.pdf} \BibitemShut
  {NoStop}%
\bibitem [{\citenamefont {Maulana}\ \emph {et~al.}(2020)\citenamefont
  {Maulana}, \citenamefont {Manna}, \citenamefont {Uykur}, \citenamefont
  {Felser}, \citenamefont {Dressel},\ and\ \citenamefont {Pronin}}]{maulana20}%
  \BibitemOpen
  \bibfield  {author} {\bibinfo {author} {\bibfnamefont {L.~Z.}\ \bibnamefont
  {Maulana}}, \bibinfo {author} {\bibfnamefont {K.}~\bibnamefont {Manna}},
  \bibinfo {author} {\bibfnamefont {E.}~\bibnamefont {Uykur}}, \bibinfo
  {author} {\bibfnamefont {C.}~\bibnamefont {Felser}}, \bibinfo {author}
  {\bibfnamefont {M.}~\bibnamefont {Dressel}}, \ and\ \bibinfo {author}
  {\bibfnamefont {A.~V.}\ \bibnamefont {Pronin}},\ }\href {\doibase
  10.1103/PhysRevResearch.2.023018} {\bibfield  {journal} {\bibinfo  {journal}
  {Phys. Rev. Research}\ }\textbf {\bibinfo {volume} {2}},\ \bibinfo {pages}
  {023018} (\bibinfo {year} {2020})}\BibitemShut {NoStop}%
\bibitem [{\citenamefont {Fang}\ \emph {et~al.}(2016)\citenamefont {Fang},
  \citenamefont {Lu}, \citenamefont {Liu},\ and\ \citenamefont {Fu}}]{fang16}%
  \BibitemOpen
  \bibfield  {author} {\bibinfo {author} {\bibfnamefont {C.}~\bibnamefont
  {Fang}}, \bibinfo {author} {\bibfnamefont {L.}~\bibnamefont {Lu}}, \bibinfo
  {author} {\bibfnamefont {J.}~\bibnamefont {Liu}}, \ and\ \bibinfo {author}
  {\bibfnamefont {L.}~\bibnamefont {Fu}},\ }\href {\doibase 10.1038/nphys3782}
  {\bibfield  {journal} {\bibinfo  {journal} {Nature Physics}\ }\textbf
  {\bibinfo {volume} {12}},\ \bibinfo {pages} {936} (\bibinfo {year}
  {2016})}\BibitemShut {NoStop}%
\end{thebibliography}%

\cleardoublepage
\section{Supplementary Information}

\subsection*{Crystal Growth and Structure Refinement}

Single crystals of RhSi were grown from the melt using Self-flux technique. Here the crystal growth was performed with an oﬀ-stoichiometric composition with slightly excess Si. First, a polycrystalline ingot was prepared using the arc melt technique with the stoichiometric mixture of Rh and Si metal pieces of 99.99 \% purity. Then the crushed powder was ﬁlled in a alumina tube and sealed inside a tantalum tube with argon atmosphere. Then the entire crystal growth was performed in a tuber furnace in argon atmosphere, kept inside a glovebox. First the sample was heated to 1500 $^\circ$C, and held there for 12 h to ensure uniform melting of the entire mixture. Then the furnace was slowly cooled to 1150 $^\circ$C with a rate of 2 $^\circ$C/h. Large single crystal chunks with average dimension of (10 $\times$ 5 $\times$ 5) mm  were obtained. The crystals were analyzed with a white beam 
backscattering Laue X-ray diffraction technique at room temperature. The samples show very sharp spots 
that can be indexed by a single pattern, revealing excellent quality of the grown crystals without any twinning or domains. 
A Laue diffraction pattern of the oriented RhSi single crystal superposed with a theoretically 
simulated pattern is presented in Fig. S1. The structural parameters were determined using a 
Rigaku AFC7 four-circle diffractometer with a Saturn 724+ CCD-detector applying graphite-monochromatized Mo-K$\alpha$ radiation. 
The crystal structure was refined to be cubic P2$_1$3 (\#198) with lattice parameter, a=4.6858(9) \r{A}.

Laue diffraction data used to orient the [100] and [010] axes within the surface along with the predicted peaks of space group 198 are shown in Fig.~S1. Measurements were performed at the Advanced Light Source at Lawrence Berkeley National Laboratory.

\subsection*{Material Symmetries}
\subsubsection*{1. Nonlinear Tensor}

The second-order optical nonlinearity generates currents at both the sum and difference frequencies of the applied electric field. LPGE and CPGE correspond to the current generated at the difference frequency,
\begin{equation}\label{eq:j}
J_i = \sigma_{ijk}E_jE_k^*
\end{equation}
For cubic space group $P2_13$ the only nonvanishing elements of $\sigma_{ijk}$ have indices $xyz$ and permutations. The elements with even permutations of $xyz$ are equal to $\sigma_{xyz}$ and odd permutations are equal to $\sigma_{xyz}^*$. If we write $\sigma_{xyz} = \gamma+i\beta$ where $\gamma$ and $\beta$ are both real, the structure of the third rank tensor can be written in the form

\begin{equation}
\sigma^{(2)} = \left(
\begin{array}{ccc}
 \left(\begin{array}{c}0 \\ 0 \\ 0 \end{array}\right)& \left(\begin{array}{c}0 \\ 0 \\ \gamma+i\beta\end{array} \right) & \left(\begin{array}{c}0  \\ \gamma-i\beta \\ 0\end{array} \right)  \\
  \left(\begin{array}{c}0 \\ 0 \\ \gamma-i\beta \end{array}\right)& \left(\begin{array}{c}0 \\ 0 \\ 0 \end{array} \right) & \left(\begin{array}{c} \gamma+i\beta  \\ 0 \\ 0\end{array} \right)  \\
  \left(\begin{array}{c}0 \\ \gamma+i\beta \\ 0 \end{array}\right)& \left(\begin{array}{c} \gamma-i\beta \\ 0 \\ 0\end{array} \right) & \left(\begin{array}{c}0  \\ 0 \\ 0\end{array} \right)  \\
\end{array}
\right)
\end{equation}
where the element $\sigma_{ijk}$ is the $k$th element of the column vector in the $i$th row and $j$th column of the outer matrix. 

\subsubsection*{2. PGEs at normal incidence}

For light incident on the [001] face, there will only be generated 2nd order effects when the electric field has nonzero components in $x$ and $y$. Eq.~\ref{eq:j} gives $J_z\propto E_xE_y$ and $J_x=J_y=0$. As currents directed in $z$ have radiation patterns that become zero in the $z$ direction, no radiation will be measured by exciting second order bulk currents at normal incidence on the [001] face.

\subsubsection*{3. Transformation properties of the CPGE}
The circular photogalvanic current can be written in terms of the photon helicity,
\begin{equation}
J_i=\beta_{ij}(\textbf{E}\times \textbf{E}^*)_j.
\end{equation}
The second rank CPGE tensor is contracted from the third-rank conductivity tensor according to the relation,
\begin{equation}
\beta_{ij}=\sigma_{ikl}\epsilon_{jkl},
\end{equation}
where $\epsilon_{jkl}$ is the unit antisymmetric tensor. Substitution of the conductivity tensor for the RhSi space group (Eq. 2) yields,
\begin{equation}
\beta_{ij}=i\beta \delta_{ij}, 
\end{equation}
where $\delta_{ij}$ is the Kronecker delta. The factor of $i$ before $\beta$ is the imaginary unit. Substitution into Eq. 3 yields,
\begin{equation}
\textbf{J}=i\beta\textbf{E}\times \textbf{E}^*,
\end{equation}
which shows that for the case of space group $P2_13$ the CPGE current is always directed parallel to the helicity vector, regardless of its direction with respect to the crystal axes. This means that CPGE polarization that depends on crystal orientation, as reported in the main text, breaks the constraints imposed by bulk symmetry.

\subsubsection*{4. Fitting LPGE data to nonlinear tensor element parameters}

Vertical and horizontal terahertz pulses are measured as a function of pump wavelength and pump linear polarization angle. For a general nonlinear LPGE tensor $\gamma_{ijk}$, illuminating a medium with linear polarization light with electric field \textbf{E} gives photogalvanic current components 

\begin{equation}
    J_i = \gamma_{ijk} E_j E_k
\end{equation}

assuming \textbf{E} has linear polarization (i.e. the components of $\textbf E$ have zero relative complex phase).
\\
\\
\indent \textbf{4.1 Pump polarization rotation}
\\
\\
In the set of LPGE experiments, we rotate the pump polarization while the sample remains fixed, and the electric field is given by $\textbf E(\theta) = E_0(\cos \theta, \sin \theta)$. This gives photogalvanic currents

\begin{equation*}
    J_x = \left( \gamma_{xxx}\cos^2\theta + \gamma_{xyy}\sin^2\theta + 2\gamma_{xxy} \cos\theta\sin\theta \right) E_0^2
\end{equation*}

\begin{equation}
    J_y = \left( \gamma_{yxx}\cos^2\theta + \gamma_{yyy}\sin^2\theta + 2\gamma_{yxy} \cos\theta\sin\theta \right) E_0^2.
\end{equation}

For each pump wavelength and terahertz polarization, the set of data has three free parameters, illustrated simply by a sine wave plus an offset with parameters $A$, $\phi$ and $C$: $A \sin(2\theta + \phi) + C$. Thus, for two terahertz polarizations, we have six free variable and can thus determine the tensor parameters $\gamma_{ijk}$ for $ijk=xxx, xxy, xyy, yxx, yxy, yyy$. Note that $\gamma_{ixy} = \gamma_{iyx}$. Data for multiple wavelengths is shown in Fig.~S2.
\\
\\
\indent \textbf{4.2 Sample rotation}
\\
\\
We can determine the same set of six parameters by keeping the pump polarization fixed at $\theta=0$ and instead rotating the sample axis about the surface normal by angle $\eta$. The two terahertz components measured will be 
\begin{multline}
    J_x(\eta) = \left(\gamma_{xxx}\cos^3\eta 
       -(\gamma_{xxy} + \gamma_{xyx} + \gamma_{yxx}) \cos^2\eta \sin\eta 
       +(\gamma_{xyy} + \gamma_{yxy} + \gamma_{yyx}) \cos\eta \sin^2\eta
       -\gamma_{yyy} \sin^3\eta\right) E_0^2
\end{multline}

and

\begin{multline}
    J_y(\eta) = \left(\gamma_{yxx}\cos^3\eta 
       +(\gamma_{xx} - \gamma_{yxy} - \gamma_{yyx}) \cos^2\eta \sin\eta 
       +(\gamma_{yyy} - \gamma_{xxy} - \gamma_{xyx}) \cos\eta \sin^2\eta
       -\gamma_{xyy} \sin^3\eta\right) E_0^2.
\end{multline}

This allows us to use two different measurement methods to determine the same set of parameters $\gamma_{iyx}$.

\subsubsection*{5. In-plane mirror symmetry}

Consider a general nonlinear tensor $\sigma$ that describes the optical response in a material such that $J_i=\sigma_{ijk}E_iE_k$. We will consider the constrains on $\sigma$ imposed by a symmetry of the mirror operation
\begin{equation}
M^x=
\begin{pmatrix}
-1 & 0\\
 0 & 1
\end{pmatrix}
\end{equation}
where, whithout loss of generality, we are restricting ourselves to 2D.

Under an operator $\mathcal{O}$, $\sigma$ will transform as

\begin{equation}
    \sigma'_{ijk}=\mathcal{O}_{i\alpha}
                  \mathcal{O}_{j\beta}
                  \mathcal{O}_{k\gamma} \sigma_{\alpha\beta\gamma}.
\end{equation}
When $\mathcal{O}$ is a symmetry of the material in question, the constraint $\sigma'_{ijk}=\sigma_{ijk}$ is imposed. For $\mathcal{O}=M^x$, we arrive at the set of equations
\begin{align}
    \sigma'_{xxx}=~~~~~(M^x_{xx})^3\sigma_{xxx}=-\sigma_{xxx}=\sigma_{xxx}~~~~(*)\nonumber\\
    \sigma'_{xxy}=(M^x_{xx})^2M^x_{yy}\sigma_{xxy}=~~\sigma_{xxy}=\sigma_{xxy}~~~~~~~~\nonumber\\
    \sigma'_{xyy}=M^x_{xx}(M^x_{yy})^2\sigma_{xyy}=-\sigma_{xyy}=\sigma_{xyy}~~~~(*)\\
    \sigma'_{yxx}=(M^x_{xx})^2M^x_{yy}\sigma_{yxx}=~~\sigma_{yxx}=\sigma_{yxx}~~~~~~~~\nonumber\\
    \sigma'_{yxy}=M^x_{xx}(M^x_{yy})^2\sigma_{yxy}=-\sigma_{yxy}=\sigma_{yxy}~~~~(*)\nonumber\\
    \sigma'_{yyy}=~~~~~~(M^x_{yy})^3\sigma_{yyy}=~~\sigma_{yyy}=\sigma_{yyy}~~~~~~~~\nonumber
\end{align}

The starred equations indicate elements that we find are equal to their own negative and therefore must be zero. We can conclude that for mirror symmetry under $M^x$, the elements $\sigma_{xxx}$, $\sigma_{xyy}$ and $\sigma_{yxy}$ must be zero.


\subsection*{AFM}

We performed atomic force microscopy (AFM) measurements over two $5\times5$~$\mu$m$^2$ regions and two $1\times1$~$\mu$m$^2$ regions (Fig.~S3). We find no patterns in the surface topography which could define any overall preferred direction in optical measurements. A surface variation of approximately 5~nm is observed.


\clearpage

\begin{figure}\centering
	\includegraphics[width=1\textwidth]{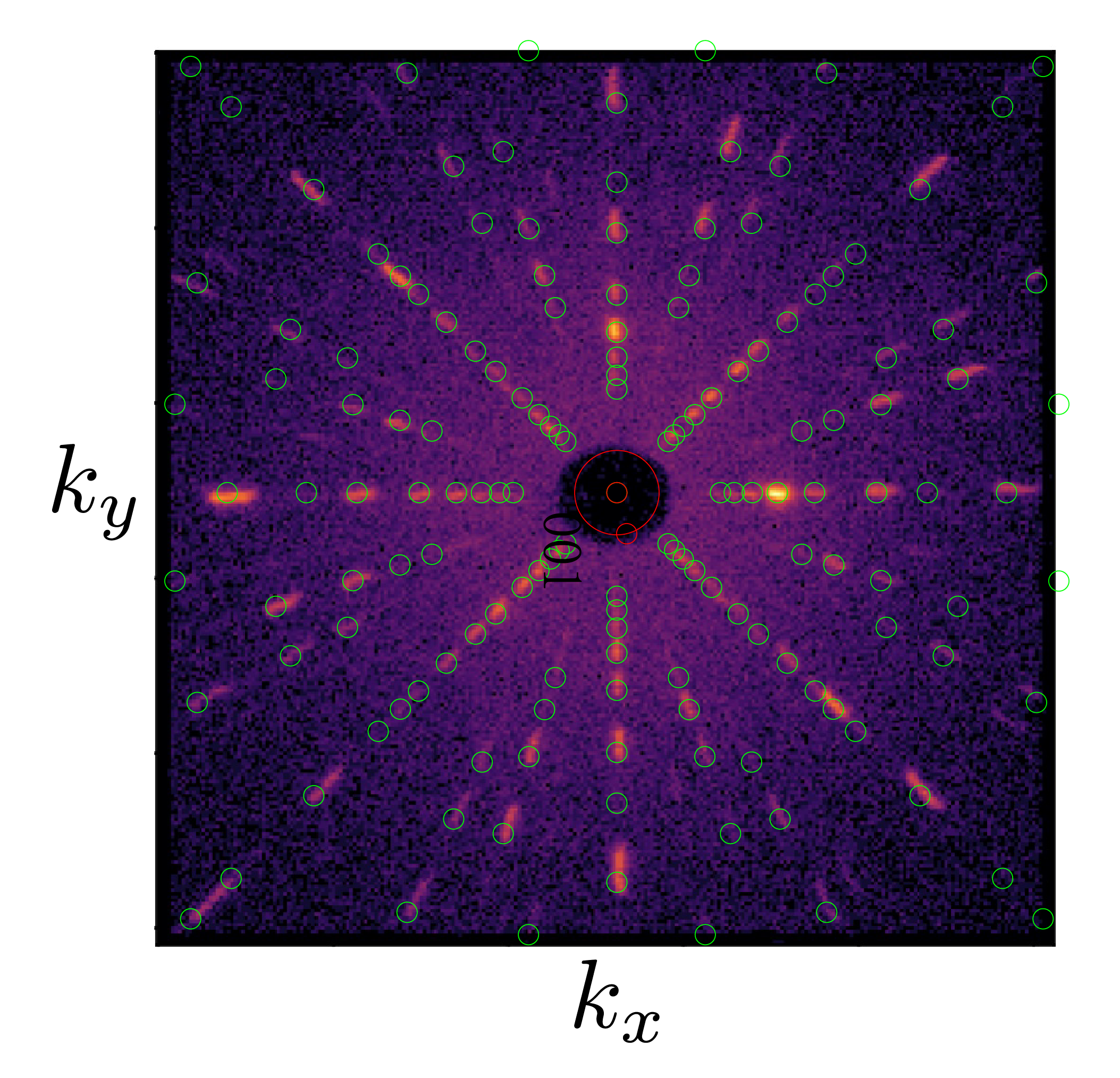}
	\caption*{\label{fig:afm}\textbf{Fig.~S1.} Laue diffraction measurement overlaid with the predicted diffraction peaks of the 001 surface of space group 198.}
\end{figure}

\clearpage

\begin{figure}\centering
	\includegraphics[width=1\textwidth]{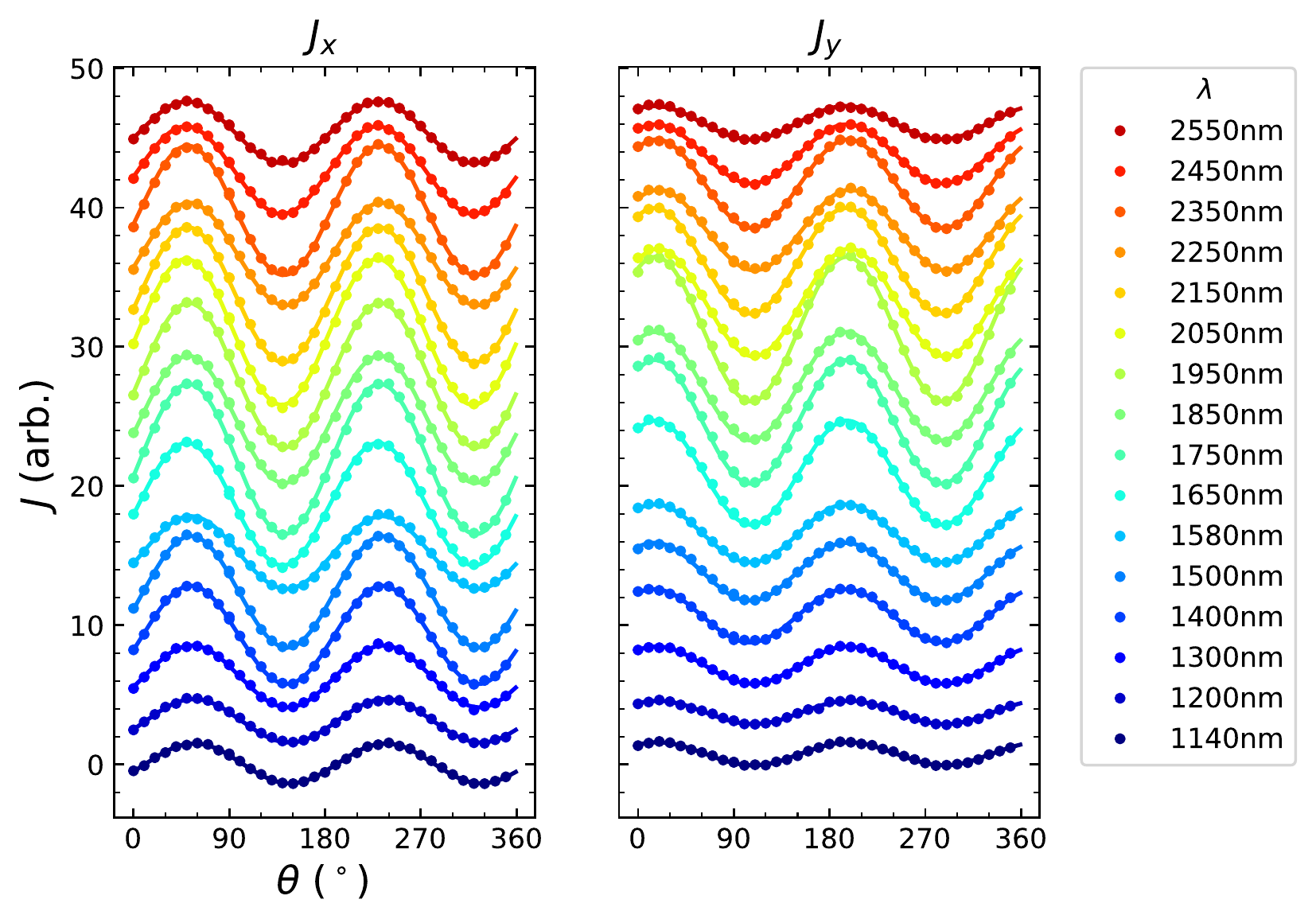}
	\caption*{\label{fig:afm}\textbf{Fig.~S2.} LPGE measurements on RhSi 001 surface as a function of pump polarization angle for $x$ and $y$ components of terahertz.}
\end{figure}

\clearpage

\begin{figure}\centering
	\includegraphics[width=1\textwidth]{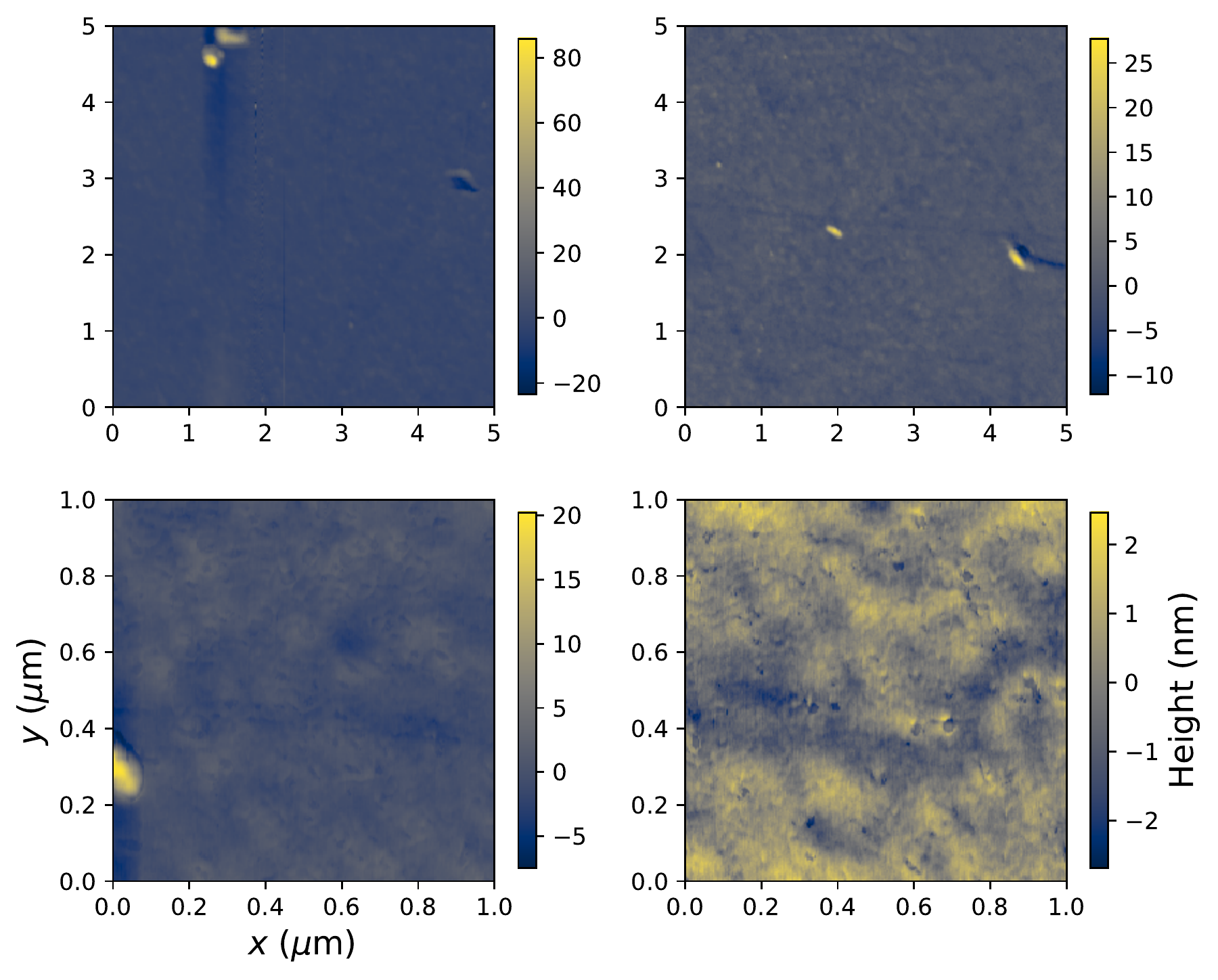}
	\caption*{\label{fig:afm}\textbf{Fig.~S3.} Four AFM measurements on RhSi [001] surface showing crystal topography.}
\end{figure}
\end{document}